\documentclass[10pt,journal]{IEEEtran}

\def\PEfrequency{400MHz}

\usepackage[utf8]{inputenc}
\usepackage{amsmath}
\usepackage{subfigure}
\usepackage{cite}
\usepackage{color}
\usepackage{siunitx}
\usepackage{booktabs}
\usepackage[table]{xcolor}
\usepackage{colortbl}
\usepackage{multirow}

\definecolor{tablecol-even}{gray}{0.97}
\definecolor{tablecol-odd}{gray}{1.0}
\newcommand{\setrowcolors}{\rowcolors{2}{tablecol-even}{tablecol-odd}}

\usepackage[noreledmac]{eledmac}

\foottwocolX{A}

\usepackage{ifpdf}
\ifpdf
  \usepackage[pdftex]{graphicx}
  \else \usepackage[dvips]{graphicx}
\fi

\renewcommand{\t}[1]{
\textnormal{#1}
}

\newcommand{\Com}[1]{
\begin{sloppypar}
\noindent{\large\bfseries\textcolor{red}{! #1}}
\end{sloppypar}
}

\title{The SpiNNaker 2 Processing Element Architecture for Hybrid Digital Neuromorphic Computing}

\author{Sebastian Höppner, Yexin Yan, Bernhard Vogginger, Chen Liu, Florian Kelber, Andreas Dixius, Stefan Scholze, Johannes Partzsch, Marco Stolba, Felix Neumärker, Georg Ellguth, Stephan Hartmann, Stefan Schiefer, Thomas Hocker, Dennis Walter, Genting Liu, Mantas Mikaitis, Jim Garside, Steve Furber, Christian Mayr
\thanks{Manuscript received \today. The research leading to these results has received funding from the European Union Seventh Framework Programme (FP7) under grant agreement no 604102 and the EUs Horizon 2020 research and innovation programme under grant agreements No 720270 and 785907 (Human Brain Project, HBP) and by the German Research Foundation (DFG, Deutsche Forschungsgemeinschaft) as part of Germany’s Excellence Strategy — EXC 2050/1 — Project ID
390696704 — Cluster of Excellence “Centre for Tactile Internet with Human-in-the-Loop” (CeTI) of Technische Universitat Dresden.
The work was supported in part by the Electronic Components and Systems for European Leadership (ECSEL) Joint Undertaking Project Technology \& Hardware for Neu-
romorphic Computing (TEMPO) under grant agreement No 826655.
The authors thank ARM and Synopsis for IP.}
\thanks{tbd are with the Faculty of Electrical and Computer Engineering, Technische Universität Dresden, Germany (e-mail: sebastian.hoeppner@tu-dresden.de) }
\thanks{tbd are with the Advanced Processor Technologies Research Group, University of Manchester}
}

\begin{document}
\bstctlcite{IEEEexample:BSTcontrol}
\maketitle

\begin{abstract}
This paper introduces the processing element architecture of the second generation SpiNNaker chip, implemented in 22nm FDSOI. On circuit level, the chip features adaptive body biasing for near-threshold operation, and dynamic voltage-and-frequency scaling driven by spiking activity. On system level, processing is centered around an ARM M4 core, similar to the processor-centric architecture of the first generation SpiNNaker. To speed operation of subtasks, we have added accelerators for numerical operations of both spiking (SNN) and rate based (deep) neural networks (DNN). PEs communicate via a dedicated, custom-designed network-on-chip. We present three benchmarks showing operation of the whole processor element on SNN, DNN and hybrid SNN/DNN networks.   
\end{abstract}

\begin{IEEEkeywords}
MPSoC, neuromorphic computing, SpiNNaker2, power management, DVFS, synfire chain
\end{IEEEkeywords}

\section{Introduction}
%
%

Neuromorphic circuits try to mimic certain aspects of neural tissue ~\cite{mead1990neuromorphic}. 
The aim is to further our understanding of how the brain computes information, 
as well as derive novel types of computational hardware for 
e.g. artificial intelligence applications ~\cite{Roy2019,mayr2019spinnaker}. 
This paper introduces the processing element (PE) architecture of the second generation SpiNNaker chip, 
a digital neuromorphic system. The "Spiking Neural Network Architecture" SpiNNaker 
is a processor platform optimized for the simulation of neural networks~\cite{furber2014spinnaker}. 
A large number of Arm processor cores is integrated in a system architecture optimized for communication 
and memory access. Specifically, to take advantage of the asynchronous, 
naturally parallel and independent sub-computations of biological neurons, 
each core simulates neurons independently and communicates via a lightweight, 
spike-optimized asynchronous communication protocol.

The second generation SpiNNaker2 scales down technology from 130nm CMOS to 22nm FDSOI CMOS 
\cite{Carter2016}, while also introducing a number of new features. 
Adaptive body biasing (ABB) in this 22nm FDSOI process node delivers cutting-edge power consumption 
\cite{Hoeppner2019a}. 
With dynamic voltage and frequency scaling, the energy consumption of the PEs scales with 
the spiking activity computed on the cores \cite{Hoeppner2017,hoeppner2019dynamic}. 
The Arm Cortex-M4 cores employed for SpiNNaker2 integrate a single-precision floating point unit, 
thus extending the fixed-point arithmetic of the first generation SpiNNaker. 
Computation-wise, SpiNNaker2 retains the processor-based flexibility of the first generation system 
\cite{Painkras2013}, while adding additional numerical accelerators to speed up common operations 
\cite{Partzsch2017b,Mikaitis2018,Neumarker2016}. In the current prototype described in this paper,
another accelerator has been added, a 16 by 4 array of 8 bit multiply-accumulate units (MAC), 
enabling faster 2D convolution and matrix-matrix multiplication \cite{kelber2020mapping}. 
Efficient usage of these accelerators can extend the simulation capacity of SpiNNaker2 
by a significant factor \cite{yan2020}. 
The PEs are arranged in groups of four to quad-processing-elements (QPEs) 
which are connected by a Network-on-Chip (NoC) to allow scaling towards 
a large neuromorphic System-on-Chip (SoC).

In the following, we introduce the processing element architecture of SpiNNaker2. 
Subsequently, we show results from the current prototype, specifically a side-by-side 
implementations of (1) a conventional spiking neural network using the numerical accelerators (2) 
a standard DNN layer using the MAC array and (3) a hybrid network, 
where we use the MAC in a spiking context.



\section{Hybrid Neuromorphic Computation Approach}
\label{sec:hybridapproach}

We try to capture our approach to neuromorphic computing with the term ''hybrid''. In a general sense, this means we aim to achieve a compromise between (a) the software/processor-based approach of Loihi \cite{Davies2018} and the first generation SpiNNaker \cite{Furber2014} and (b) the classical neuromorphic approaches with dedicated neuron and synapse circuits (in digital or analog) \cite{merolla2014million,frenkel2018,moradi2017scalable,mayr2016}. Thus, while we do not implement the full neuron and synapse functionality as dedicated circuits, we aim to achieve close to this kind of efficiency by numerically accelerating common operations such as exponentials, log or random functions. At the same time, classical neuromorphic approaches have always been hindered by their essential hard-coding of functionality, which we aim to avoid by keeping the processor cores at the center, employing them to tie the accelerator blocks together. 

In an extension of the above definition, ''hybrid'' also means that we can realize both spiking and rate-based artificial neural networks (SNN/ANN) on SpiNNaker2, since we incorporated accelerators for both types of networks and the ARM processors and communication infrastructure (Network on chip, DMA, etc) are flexible enough to support both. Thus, hybrid networks at the boundary between SNNs and ANN can be realized that combine the best of both worlds, e.g. the numerical simplicity of ANNs with the sparsity and time-dynamic features of SNNs, for enhanced energy efficiency and novel computational paradigms. In a specific instantiation of this idea, the MAC array could be run not frame-based, but in an event-triggered fashion. To fully use the multipliers, this mode of operation needs a graded weight (usually given) and a graded activity-related input value (usually not given, if we assume single-bit axonal spikes). Biologically realistic versions of a graded input value (i.e. multi-bit spikes) for the multipliers in the MAC array could be (1) electrical synapses that generally transmit graded values, e.g. in a retina model. (2) Dendritic computation, i.e. where amplitudes inside a neuron are scalar and have some amplification factor (weight) when being transmitted to other parts of the neuron. Or (3) treating the neuron input as a scalar synaptic current (e.g. representing a pixel grayscale value). This approach is employed in our hybrid use case later in the paper. Non-biologically realistic versions would use the possibility of a `spike with a payload' for e.g. more efficient coding by trading spike payload for spike frequency, achieving increased sparsity \cite{voelker2020spike}. Another possibility could be more complex processing. That is, the payload could represent some dynamic characteristic of the neuron, e.g. the value of some adaptation process, or how fast the charging slope was, etc.

\section{SpiNNaker 2 Many-Core Architecture}
\label{sec:qpe}
With the shrink of semiconductor technologies more processing elements (PEs) can be integrated on a single silicon die,
which improves scalability toward multi-million core neuromorphic processors systems. 
However, the on-chip bus system (system NoC) from SpiNNaker \cite{Painkras2013} does not scale toward a larger number of on-chip PEs,
since it is implemented with flat logic on the chip top-level. 
This approach is not feasible for a larger number of PEs since flat top-level routing and timing closure would not be feasible. 
Therefore a quad-processing element (QPE) architecture is employed here, as described in the following.

\subsection{QPE and NoC Architecture}

\begin{figure}[htb]
	\centering
		\includegraphics[width=0.47\textwidth]{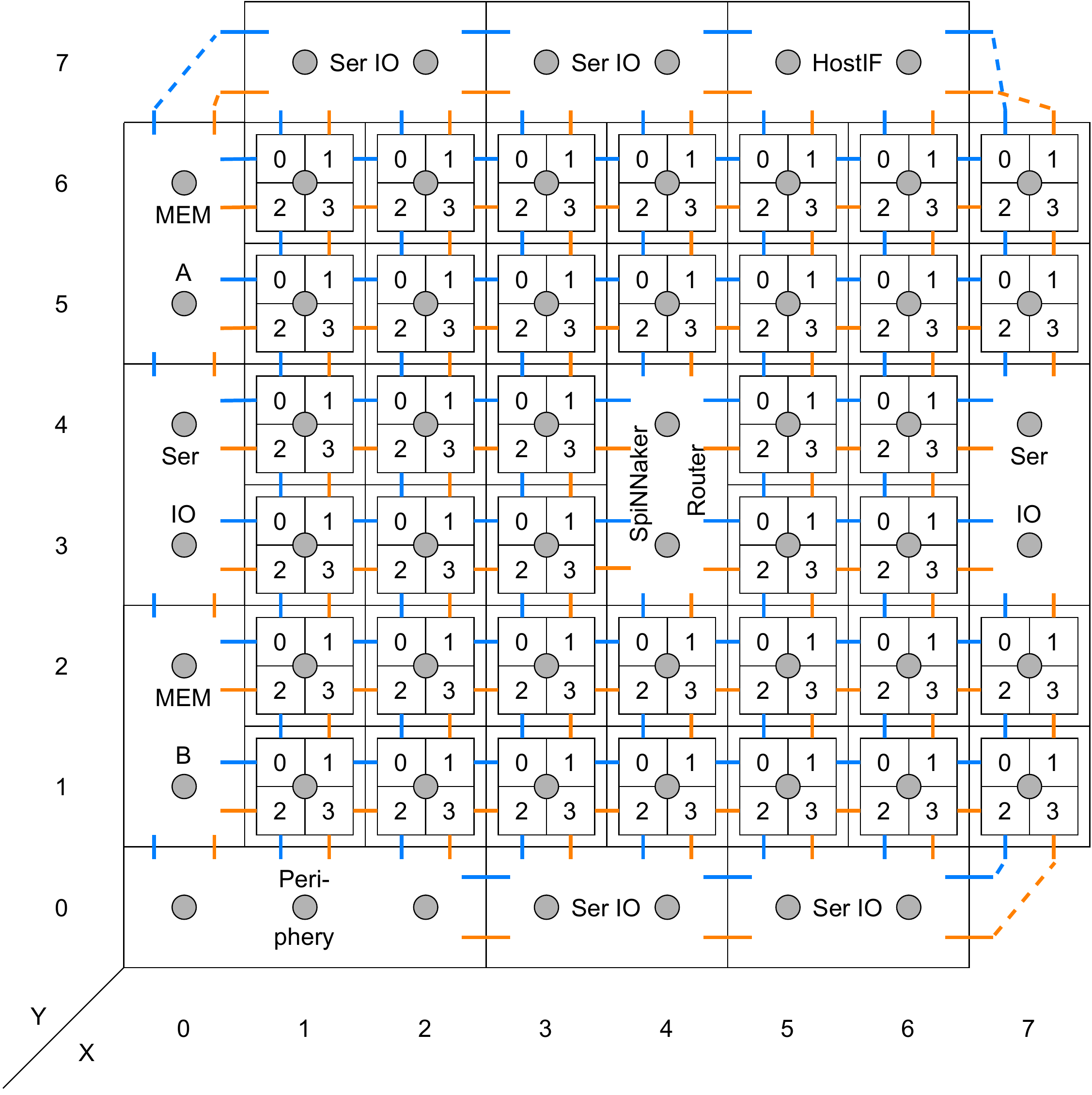}
	\caption{SpiNNaker2 NoC Topology and top-level Floorplan, DNOC in blue and CNOC in orange}
	\label{fig:SpinNNaker2_floorplan}
\end{figure}

\begin{figure}[htb] 
	\centering
		\includegraphics[width=0.47\textwidth]{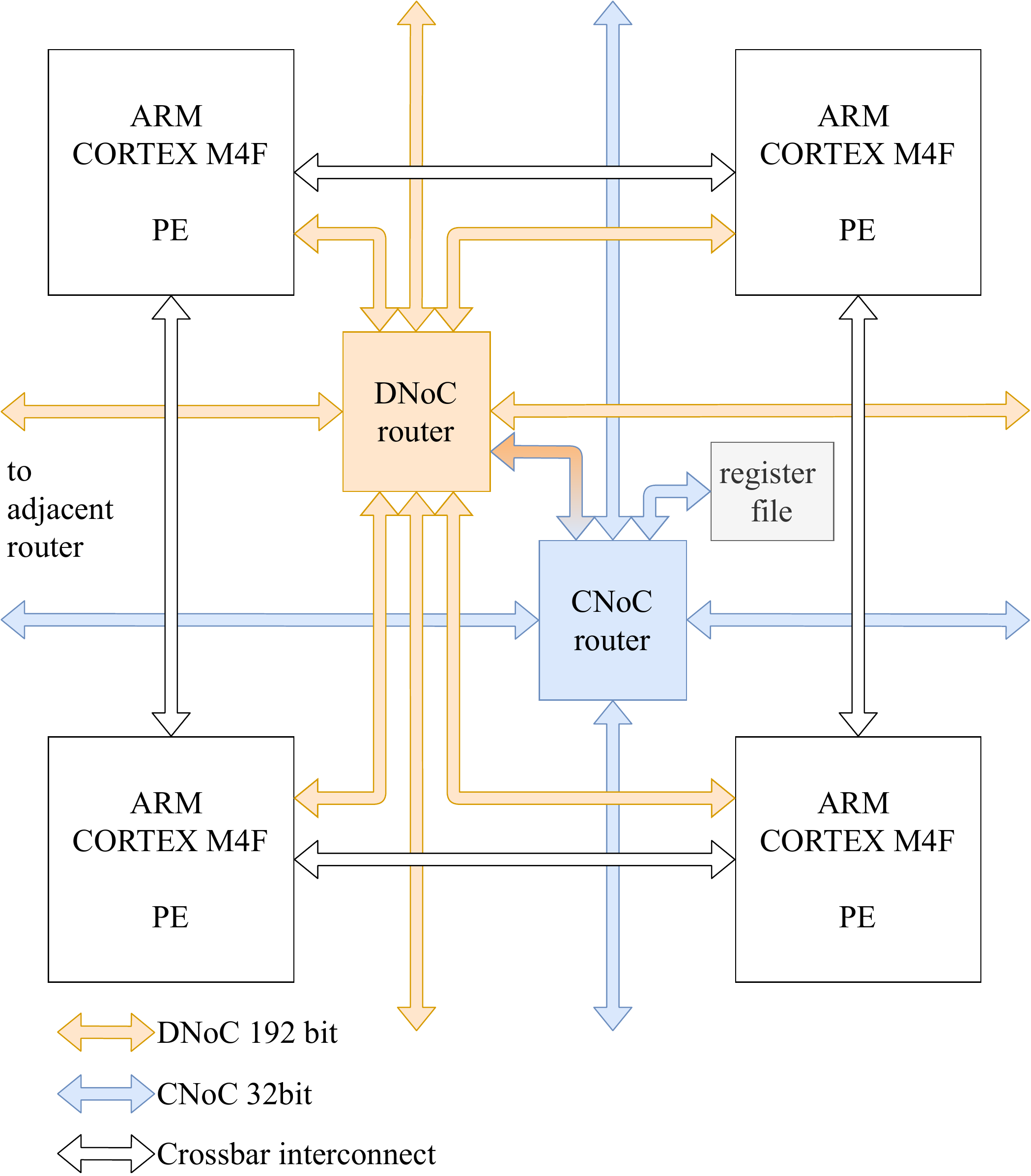}
	\caption{QPE NoC architecture}
	\label{fig:QPE-NoC}
\end{figure}


Fig.~\ref{fig:SpinNNaker2_floorplan} shows the top-level floorplan architecture of SpiNNaker 2.
Its main components are 38 QPEs with 152 PEs in total. 
The SpiNNaker 2 packet router (Sec.~\ref{sec:spinnakerrouter}) is responsible for spike communication. 
In total 6 chip-to-chip communication links allow to form the SpiNNaker multichip network \cite{Painkras2013}.
A host interface allows to connect to an Ethernet infrastructure. 
Two LPDDR4 memory controllers and PHYs allow the connection of off-chip DRAM. 
A periphery block contains various low-speed interfaces (GPIO, JTAG, SPI) for configuration, debugging and other general purpose use cases.

\begin{figure}[htb]
	\centering
		\includegraphics[width=0.47\textwidth]{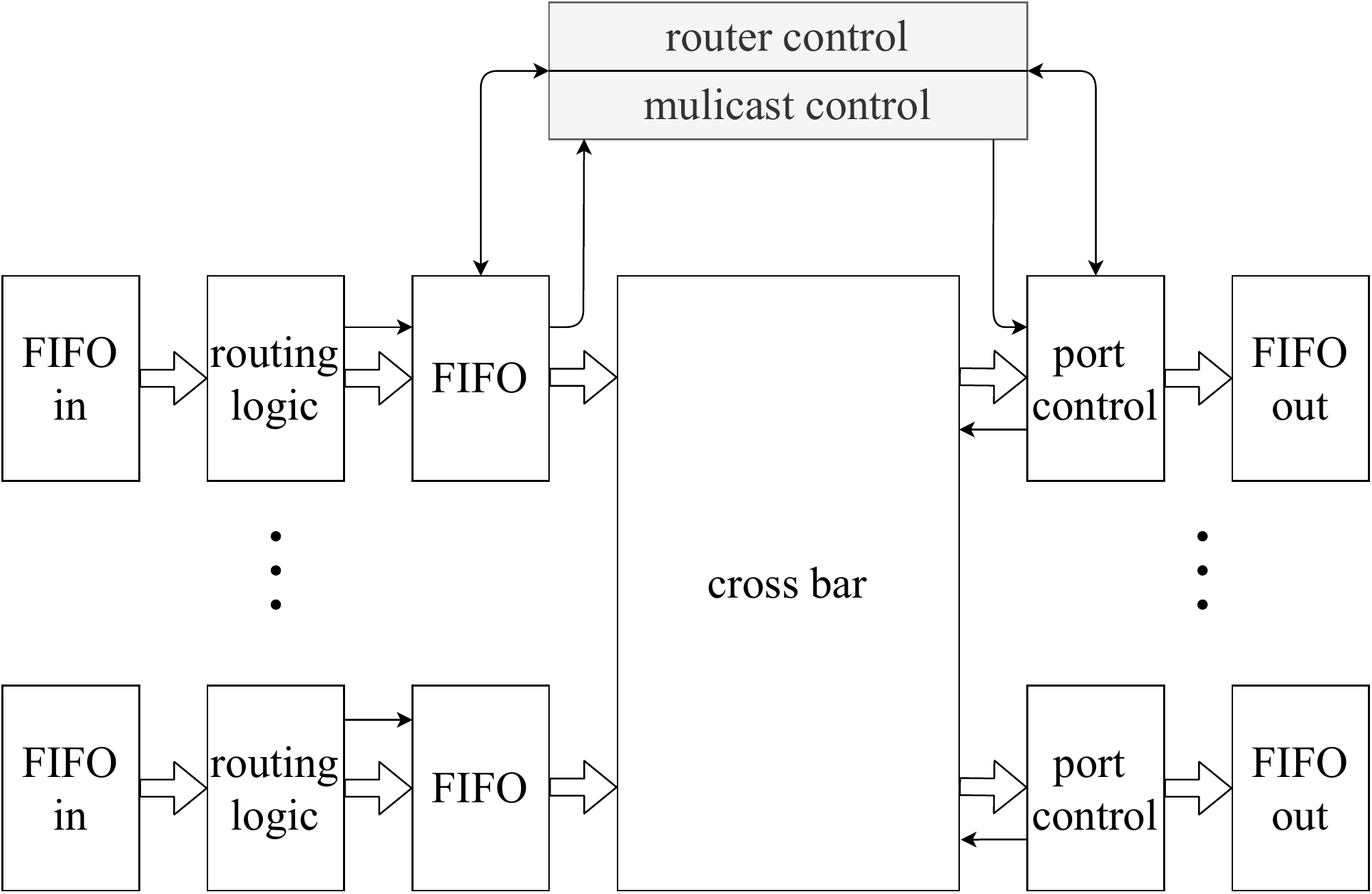}
	\caption{NoC router architecture}
	\label{fig:NoC_router}
\end{figure}

\begin{figure}[htb]
	\centering
		\includegraphics[width=0.47\textwidth]{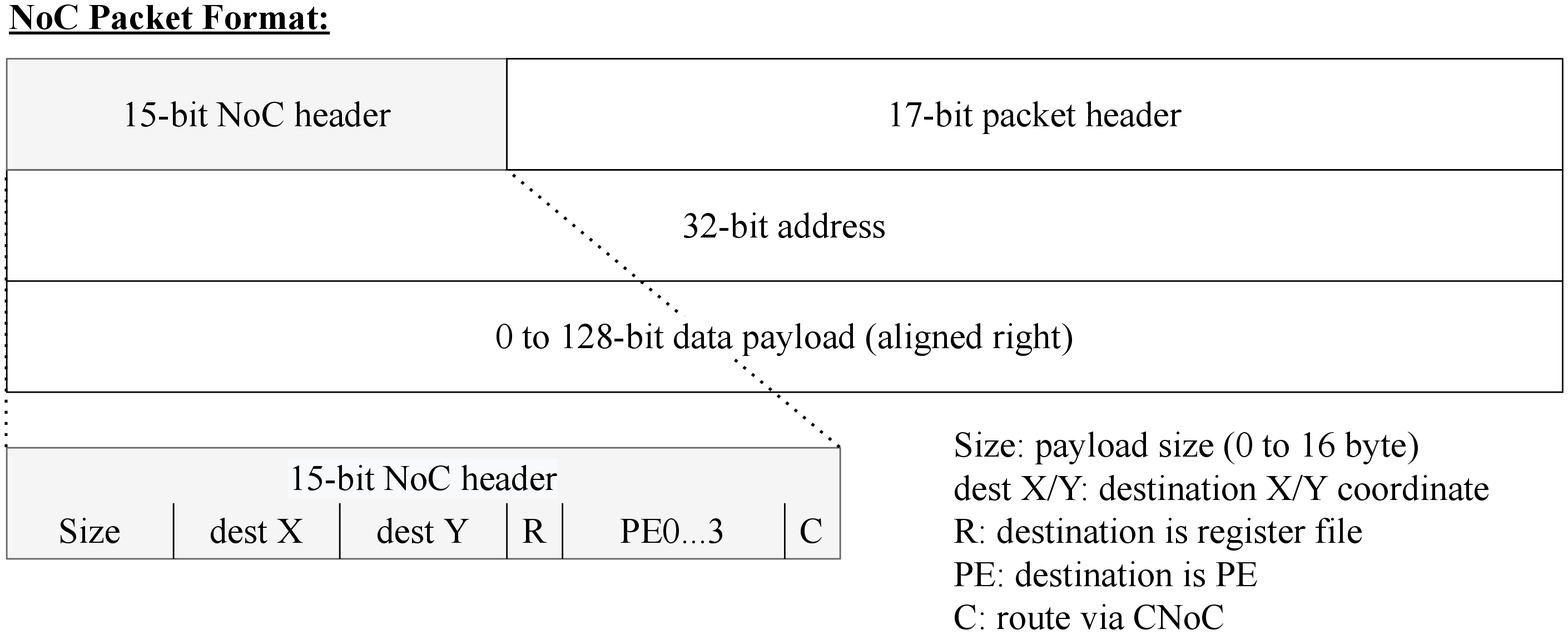}
	\caption{NoC packet format}
	\label{fig:NoC_packet_format}
\end{figure}

A detailed view of the QPE is shown in Fig.~\ref{fig:QPE-NoC}. 
The QPE logically combines 4 PEs and a NoC router, with connections to the PEs and the four neighbor QPEs. 
It is implemented in a globally-asynchronous-locally-synchronous (GALS) clocking scheme, 
allowing the PEs to operate in the dynamic voltage and frequency scaling (DVFS) scheme from \cite{hoeppner2019dynamic} 
independent from each other and the NoC router logic. Additionally, 
the GALS approach also decouple the QPEs from each other to prevent the need for a chip top-level synchronous clock distribution network.
Therefore, the QPE further forms a place and route hard macro which can be seamlessly arranged in an array on chip top-level to build a the many-core SoC.
All signal connections to neighbor QPEs are realized in a connect-by-place scheme with asynchronous FIFO clock domain crossings.
Additionally ot the NoC connections, only a small number of dedicated signals are routed between the QPEs, 
as for example clock, reset and a few physical interrupt signal lines.
The NoC is the main communication resource that interconnects all on-chip components in a 2D-mesh structure. 
It consists of two independent but interconnected NoC meshes: the data NoC (DNoC) and  configuration NoC (CNoC).
For data transfers such as DMA or spike packets the DNoC is used. 
A simplified structure of the NoC router is shown in Fig. ~\ref{fig:NoC_router}. 
It consists of wide asynchronous FIFOs for synchronous incoming packets to the router's clock domain. 
The packet destination is determined by a X/Y-first routing logic. 
A following FIFO stage is shorten the internal critical timing path. 
The output port control arbitrates incoming routing requests in a round-robin fashion and controls the crossbar accordingly. 
DNoC flit size is 192bit hence an entire packet can transferred at once for high throughput. 
The router's latency is 5 clock cycles per hop at a frequency of \PEfrequency. 
Besides the NoC a separated CNoC is implemented.
It is operated using the reference clock signal and is thereby operational if no clock generator is switched on. 
And provides access to all registers in a boot-up and configuration step. 
During operation it allows for data transmission, even if the DNoC is blocked, e.g. by large DMAs.
It has a flit width of 32bit and is operating in a worm-hole switching scheme. 
For interoperability DNoC and CNoC are using the identical packet format shown in ~\ref{fig:NoC_packet_format}.
The 192bit NoC packet consists of a 15bit NoC header, 17bit packet header, 32bit address and 0 up-to 128bit payload data.
Routing decisions of the NoC are made based on the X/Y destination coordinates of the NoC packet. 
Additionally there are four destination PE bits reserved to allow a multicast on QPE level for spike packet or data packets.

Within SpiNNaker2 mainly data packets and SPiNNaker packets (see Sec.~\ref{sec:spinnakerrouter}) are transmitted. 
The NoC allows to abstract communication to the PEs also for infrastructure signals. 
This includes interrupt packets or dedicated test data streams for scan tests of the PEs.

\subsection{The SpiNNaker Router}
\label{sec:spinnakerrouter}
\begin{figure}[htb]
	\centering
		\includegraphics[width=0.47\textwidth]{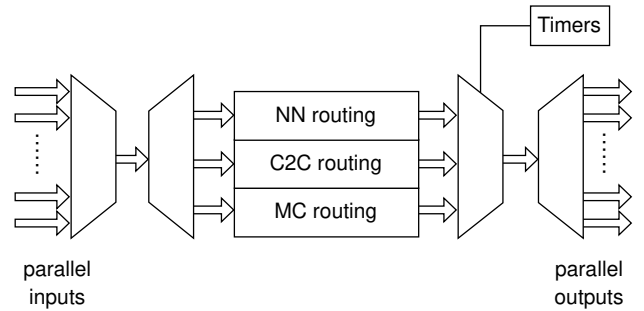}
	\caption{SpiNNaker router}
	\label{fig:SPRouter}
\end{figure}

\begin{figure}[htb]
	\centering
		\includegraphics[width=0.47\textwidth]{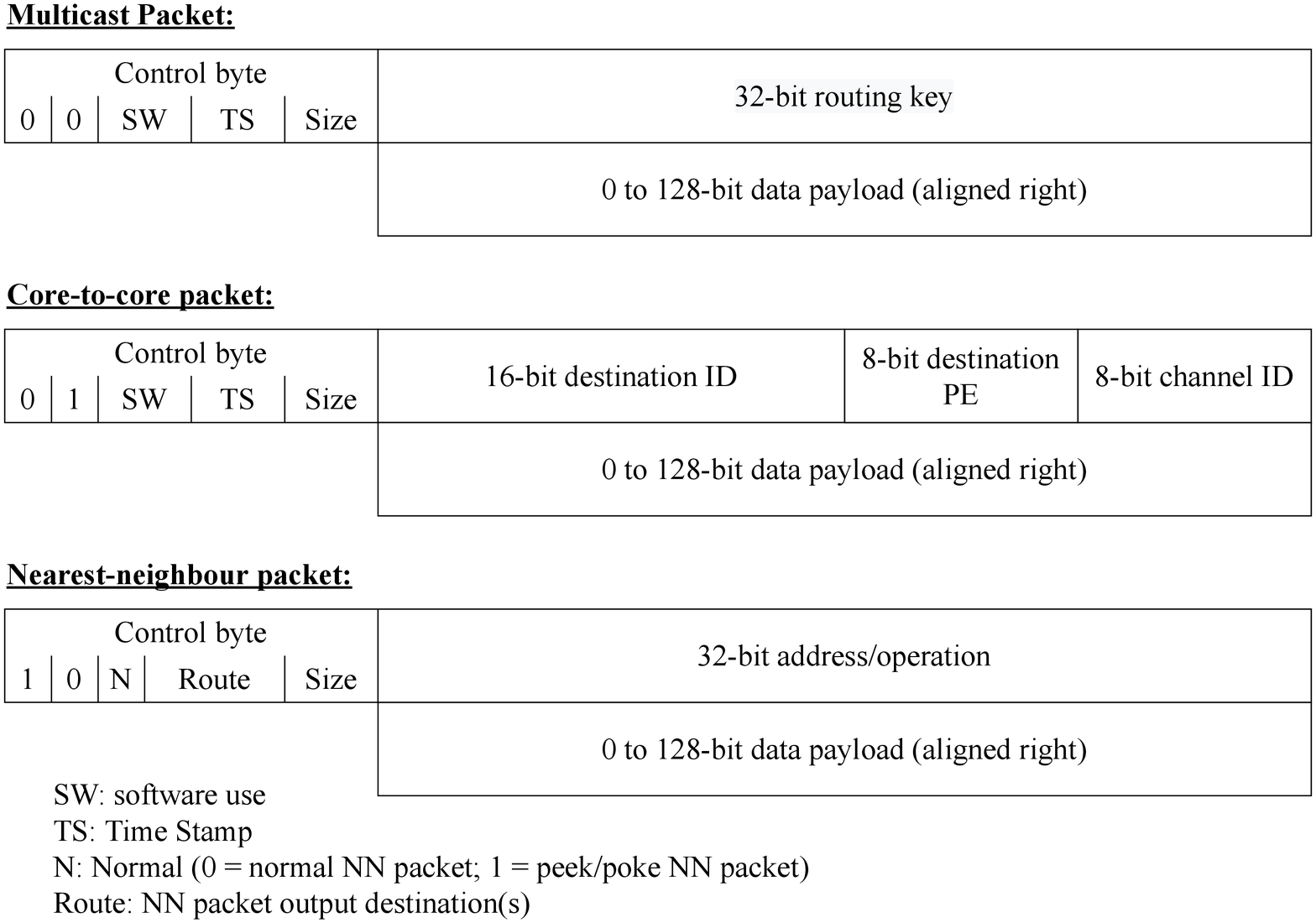}
	\caption{SpiNNaker packet format}
	\label{fig:spkt2_format}
\end{figure}

The SpiNNaker2 packet router, based on \cite{Wu2009,Wu2010}, is one of the key components of the chip. 
It is responsible for routing multicast, core-to-core and nearest neighbour packets. 
An overview of the SpiNNaker2 packet router architecture is shown in Fig.~\ref{fig:SPRouter}. 
The router incorporates a number of evolutions to support the increased communication throughput, 
such as parallel routing structure, larger multicast routing tables, improved packet-dropping mechanism, 
out-of-order issue buffer and a large optional packet payload. 
Error Correction Code (ECC) SRAMs are adopted and TCAM built-in self-test is designed to improve fault tolerance 
and test automation. Additionally, the router has fully payload pipeline control and a clock gating implementation for power-saving. 
The multicast packets are used for neural events and routed by a key provided at the source. 
The core-to-core packets are routed by a destination address to any core on any chip and intended for machine management. 
Last but not least, the nearest neighbour packets are routed by destination port(s) to the monitor processor(s) of the neighbouring chip(s), 
and they are intended for machine boot and debugging functions. The packet format is shown in Fig.~\ref{fig:spkt2_format}.

\subsection{Processing Element}
\begin{figure}[htb]
	\centering
		\includegraphics[width=0.47\textwidth]{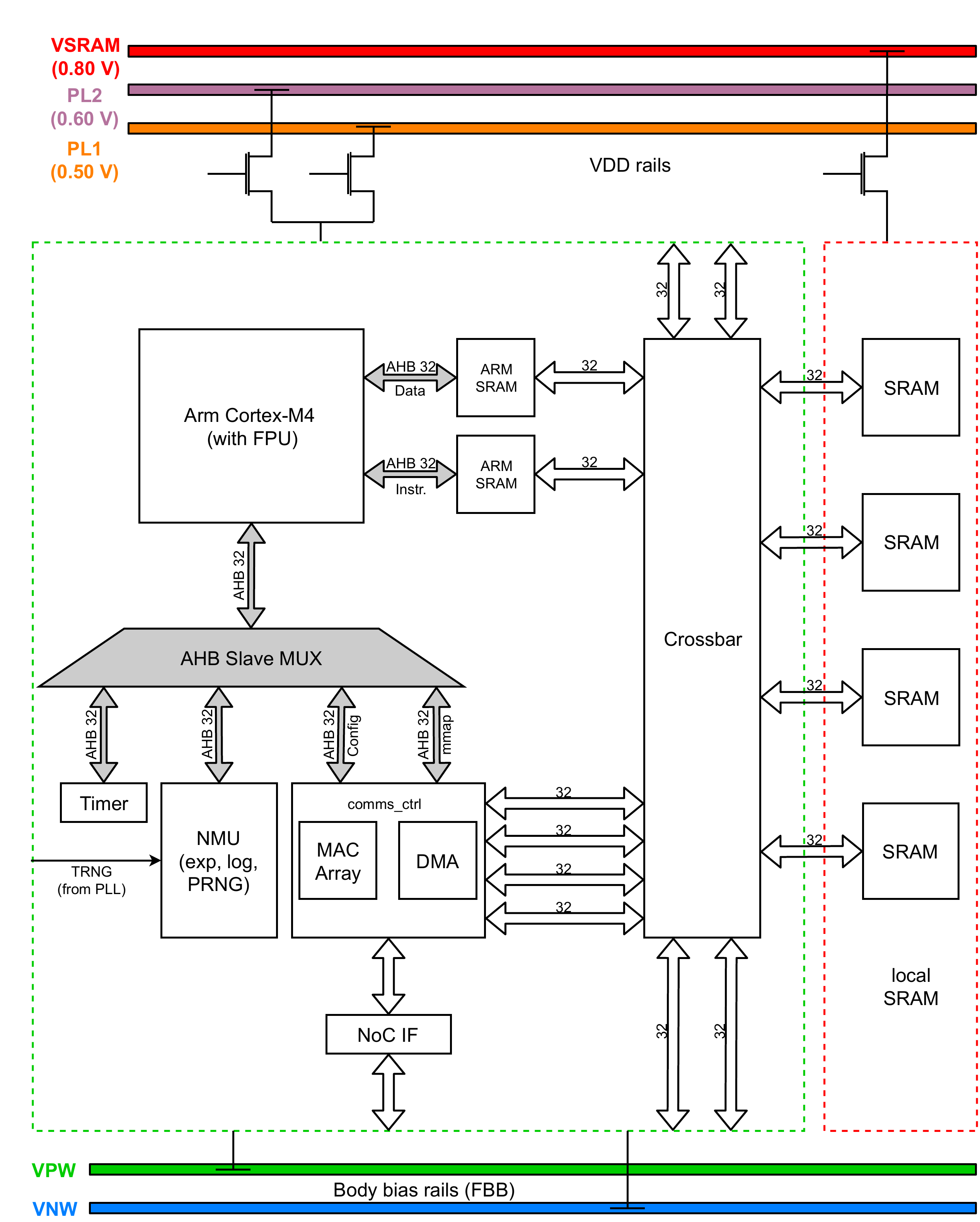}
	\caption{PE power management architecture}
	\label{fig:PE_PM_achitecture}
\end{figure}

As shown in Fig. \ref{fig:PE_PM_achitecture}, the processing Element (PE) consists of the Arm Cortex-M4F core, 
pseudo and true random number generators \cite{Neumarker2016}, 
a fixed-point exponential and logarithm accelerator \cite{Partzsch2017b,Mikaitis2018}, 
a MAC array, timers, SRAM memory array, AHB bus, DMA, and crossbar. 
DVFS \cite{hoeppner2019dynamic} allows for switching between two VDD rails (PL1 and PL2 in Fig. \ref{fig:PE_PM_achitecture}) during operation of the PE.

While the Arm core forms the main computational resource of the PE, 
the additional accelerators like the random number generators and 
the MAC array can greatly increase the computational and power efficiency for certain applications \cite{synsampling19}\cite{yan2020}.
To reduce the possibility of contention for SRAM, the SRAM is divided into four addressable banks.

The various computation and memory units of the PE are interconnected through the communication units AHB bus, DMA and crossbar. 
As shown in Fig. \ref{fig:PE_PM_achitecture}, the Arm core is master of three AHB buses, two for access to the memory, 
one for access to the accelerators, DMA and timers. 
In addition to the AHB buses for data and instruction for the Arm core of the same PE,
the crossbar is also connected to neighboring PEs withing the same QPE, 
which allows low latency memory sharing between PEs in the same QPE. 
The crossbar is also connected to the DMA for high-speed communication with other PEs or the DRAM, 
and for the MAC array to read and write data independently of the the Arm core.

The different numerical accelerators are each connected as individual slave to an AHB multiplexer. 
Each accelerator has a specific range in the memory map of the PE that it uses for read and write.
For details on specific numerical accelerators, please refer to the respective publications, 
\cite{Mikaitis2018} for the exp/log accelerator, \cite{Neumarker2016} for the true random number generator.


\begin{figure}[htb]
	\centering
		\includegraphics[width=0.47\textwidth]{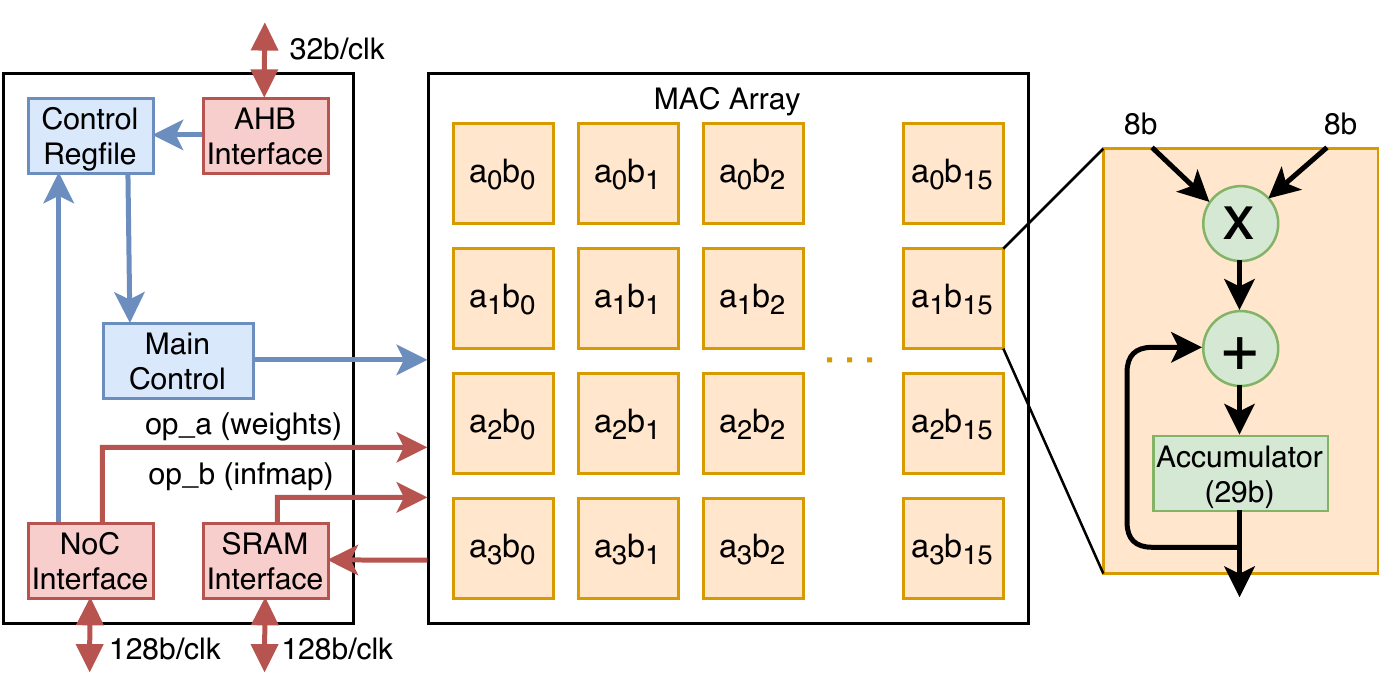}
	\caption{MAC accelerator structure}
	\label{fig:MLA}
\end{figure}

With this testchip iteration the SpiNNaker2 system introduces a broadcast output-stationary
MAC  accelerator shown in Fig. \ref{fig:MLA}, targeting core operations in popular Deep Neural Network models.
The first revision provides 8bit unsigned 4x16 multiply-accumulate per clk cycle
and two modes of operation, 2D convolution (CONV) or matrix multiplication (MM). 
These modes are realized by  adapting the memory fetch pattern to the MAC array. To not get bottlenecked by the memory fetch,
the accelerator almost fully utilizes the 128Bit/clk local connection to the SRAM within the same PE and 
writes out results via the same connection. 
The second operand is fetched from the NoC via a 128Bit/clk NoC interface, which is only partially utilized.
The accelerator can be controlled over either the ARM core or per incoming
configuration NoC packets. Once started, its execution is independent of the ARM core and from other processing elements. 
An interrupt is thrown if an operation has finished. 
For CONV operation, a shift register is included for input feature map reuse relaxing the continous memory fetch to 4Byte/4clk.
Furthermore the output channel dimension and the input feature map width dimension
were chosen to fully utilize 1x1 kernels of bottleneck layers and exploit weight and
input feature map reuse. The reduced size of the array
was chosen due to area considerations. However, since each PE includes its own independent
MAC array, the system supports highly distributed inference operations. For that,
distribution strategies to optimize for time efficiency and data reuse were developed and
simulated \cite{kelber2020mapping}. This also enables row-stationary dataflow 
\cite{journals/micro/ChenES17} and further data reuse due to weight and input feature map sharing between PEs.
Further details on how the accelerator works in more detail can be seen in \cite{zeinolabedin202216}.

\section{22FDX Implementation}
\label{sec:22FDX}
\subsection{Adaptive Body Biasing}
Adaptive body biasing (ABB) is a technique for FDSOI technologies \cite{Carter2016} 
for the compensation of device performance variations caused by process, voltage and temperature (PVT) variations 
by means of the adaptive control of the back-gate voltages \cite{Hoeppner2019a}.
In the 22nm FDSOI target technology \cite{Carter2016}, two body bias schemes, Forward-Body-Bias (FBB) and Reverse-Body-Bias (RBB), 
are available. In this work the FBB scheme with the flipped well approach (N-Well below NMOS, P-Well below PMOS) is used. 
Compared to the zero-bias approach where 0.0V is statically connected to the biased-wells, 
an up to 10X speed improvement at 0.50V operation can be achieved \cite{Hoeppner2019a}. 
This enables the ultra-low-voltage (ULV) implementation of the PEs at reasonable clock frequencies of several 100MHz. 

\begin{figure}[htb]
	\centering
		\includegraphics[width=0.47\textwidth]{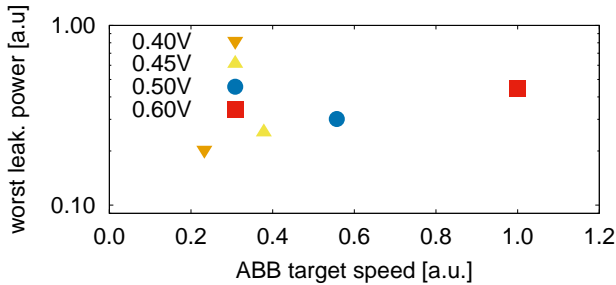}
	\caption{Relative performance and leakage power of ABB biased designs. Y axis shows worst case leakage of example cells (SLVT28) in ffg VDD+10\% 125C corer. X-axis shows the target speed of the ABB regulation}
	\label{fig:abb_lib_performances}
\end{figure}

In this work the ABB IP platform \cite{Hoeppner2019a} is used, 
which provides ABB-aware characterized libraries and dual-rail SRAM from 0.40V to 0.80V core supply voltages. 
Design implementation, sign-off and power analysis has been performed following the ABB-aware methodology from \cite{Hoeppner2019b}. 
Fig.~\ref{fig:abb_lib_performances} shows the relative performance and 
leakage power of the nominal supply conditions from 0.40V to 0.60V, being considered for this work. 
It shows the target speed for ABB regulation, as performance indicator for the operating frequency, 
versus the worst-case leakage power of the Super-Low-Vt (SLVT) flavor with 28nm gate length at the fast-hot PVT condition. 
From 0.40V to 0.60V there is a performance gain by factor 5 with a worst case leakage increase by factor 2. 

\subsection{Design implementation Strategy}
\begin{figure}[htb]
	\centering
		\includegraphics[width=0.47\textwidth]{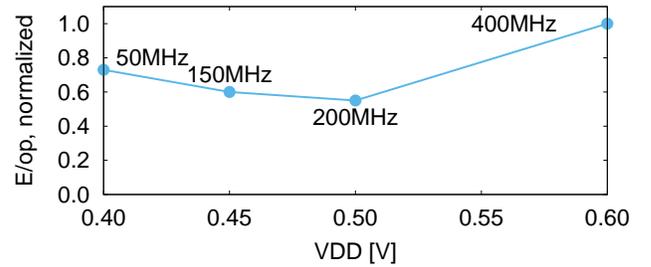}
	\caption{Trial implementation result, relative energy consumption from design implementations at 4 different supply voltages and target clock frequencies, MEP Study result}
	\label{fig:PE_MEP_study}
\end{figure}

The main implementation constraint of the PE was to achieve minimum energy per operation at a reasonable clock frequency, 
similar to the first generation SpiNNaker chip \cite{Painkras2013} with a processor clock frequency of 180MHz. 
Therefore, a mixture of Super-Low-Vt (SLVT) and Low-Vt (LVT) standard cells from a 9-track 104nm CPP library 
is used \cite{Hoeppner2019b} to achieve the speed target under ULV conditions. 
Trial implementations of the PE have been performed at 4 different nominal voltages and different target clock frequencies, 
as shown in Fig.~\ref{fig:PE_MEP_study}. 
The designs have been implemented for timing closure at the worst speed PVT corner (ssg, VDD -10\%, -40C) and 
power analysis has been performed in the worst case power corner (ffg, VDD+10\%, 125C) 
with on post-layout VCD based power analysis of the processor running a software task. 
Note that the absolute worst speed considered for sign-off does not scale similar to the nominal 
target speed as plotted in Fig.~\ref{fig:abb_lib_performances} since additional margins apply \cite{Hoeppner2019b},
which lead to significant additional performance degradation especially at the extreme low voltage conditions. 
Considering energy per operation metric, there exists a minimum energy point (MEP) at nominal 0.50V operation. 
At higher voltages, more energy per operation is spent due to higher switching energy. 
At lower voltages more energy per operation is accumulated due to leakage power over the longer clock period.  

\begin{figure}[htb]
	\centering
		\subfigure[standard cell distribution \label{fig:pe_impl_cell}]{\includegraphics[width=0.47\textwidth]{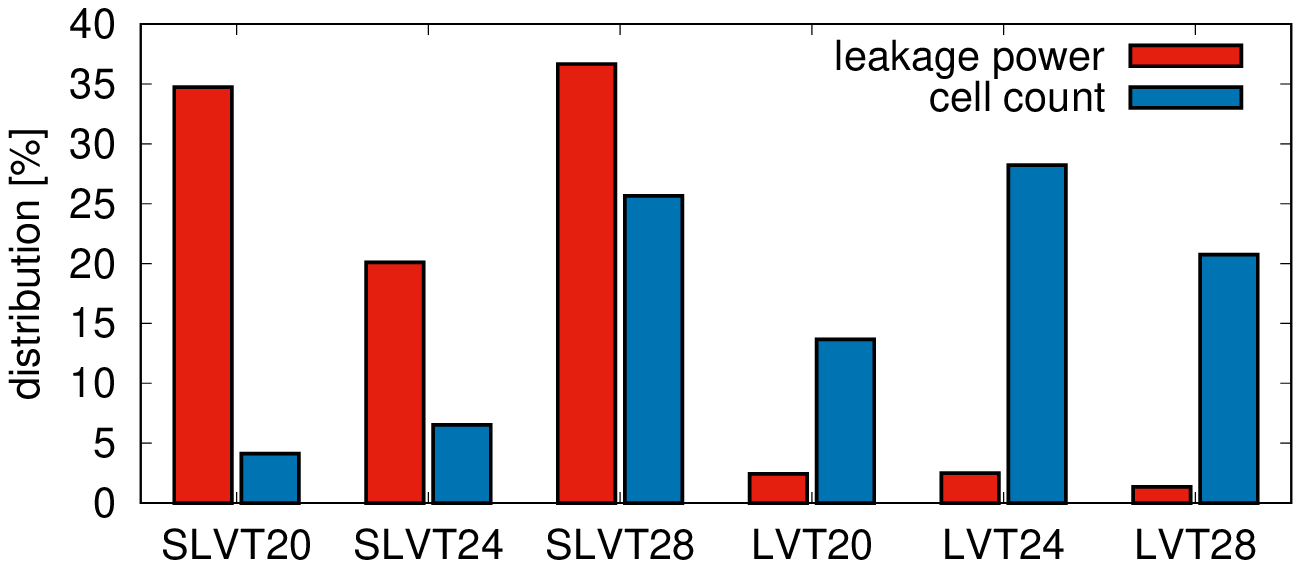}}
		\subfigure[component distribution \label{fig:pe_impl_comp}]{\includegraphics[width=0.47\textwidth]{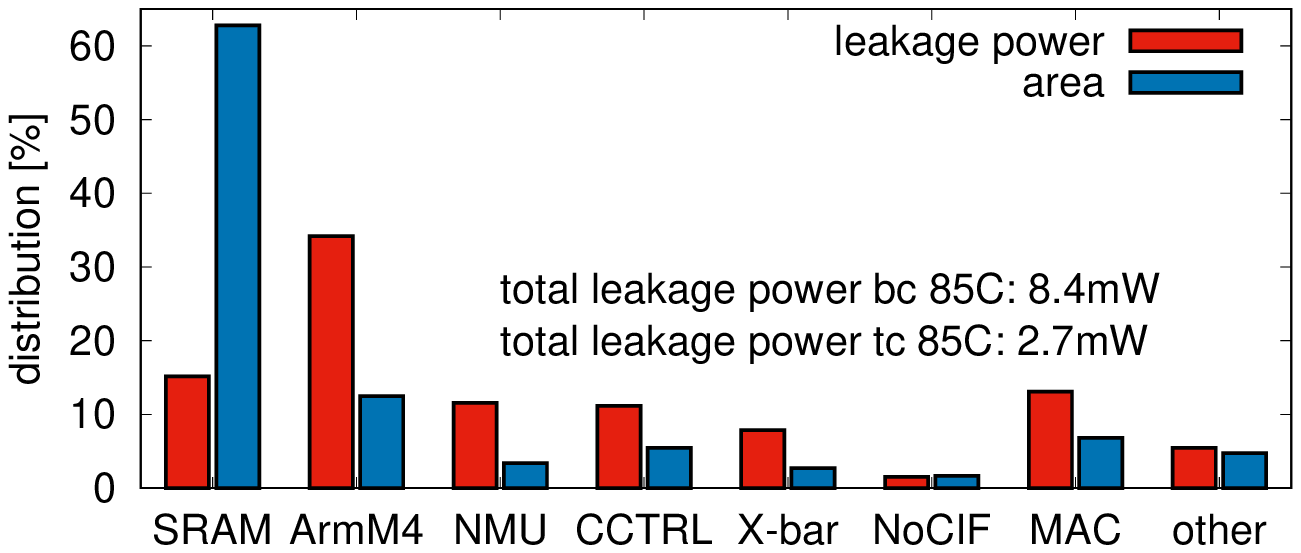}}
		\caption{PE implementation results statistics}
	\label{fig:pe_implementation_statistics}
\end{figure}

As result, the target implementation point has been chose at 0.50V nominal and 200MHz. 
However, this does not denote a significant performance scaling compared to the first generation SpiNNaker processor \cite{Painkras2013}. 
Therefore, the DVFS technique from \cite{Hoeppner2017,hoeppner2019dynamic} is applied here, with two performance levels (PLs). 
PL1 is the MEP operating point of (0.50V, 200MHz) and PL2 is defined as the higher performance level at (0.60V, 400MHz). 
This allows for DVFS operation between 200MHz and 400MHz. 
Fig.~\ref{fig:pe_implementation_statistics} summarized the implementation results statistics of the PE for cell distribution and
component area and leakage distribution. 
By means of ABB the design could be implemented with  relatively low SLVT cell-count (Fig.~\ref{fig:pe_impl_cell}) 
which still dominates the leakage power. 
The total PE cell area is dominated by the 128kB SRAM and the Arm Cortex-M4 MCU with floating point unit 
which in sum consume 75\% of the PE area. The peripheral blocks like hardware accelerators, 
NoC and bus interfaces consume only 25\% of the area (Fig.~\ref{fig:pe_impl_comp}), 
but 50\% of the total leakage.  


In Fig.~\ref{fig:jib1_chipphoto} the layout of the QPE with 4PEs is shown. Because the QPE logic shows less toggle rate compared to the PE its energy efficiency is more sensitive to leakage and the NoC to achieve high peak throughput to handle the traffic from the PEs running at PL2, a nominal 0.60V implementation at 400MHz has been chosen for the QPE logic, including the NoC router and FIFOs.


\section{Testchip}
\label{sec:testchip}
\begin{figure}[htb]
	\centering
		\includegraphics[width=0.35\textwidth]{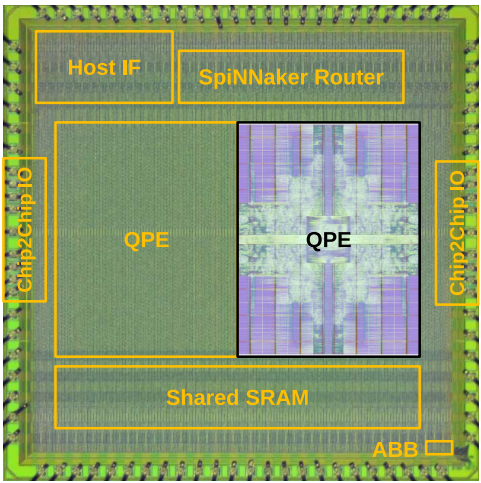}
	\caption{Chip photo and QPE layout}
	\label{fig:jib1_chipphoto}
\end{figure}

A test chip JIB has been implemented in GLOBALFOUNDRIES 22FDX technology \cite{Carter2016} 
with a chip area of 8.76$\t{mm}^2$. 
The chip photo is shown in Fig.~\ref{fig:jib1_chipphoto}.
It contains two QPEs with 8 PEs in total, a SpiNNaker router instance for Spike communication 
with two prototype chip-2-chip connection SerDes links.
A SerDes host interface is integrated to allow for connection of an FPGA. 
The periphery block contains various standard interfaces (UART, SPI, I2C, JTAG) and 
the instance of the adaptive body bias generator \cite{Hoeppner2019b}. 
The on chip clocks are generated using the clock generator from \cite{Schraut2019}.  

\begin{figure}[htb]
	\centering
		\includegraphics[width=0.35\textwidth]{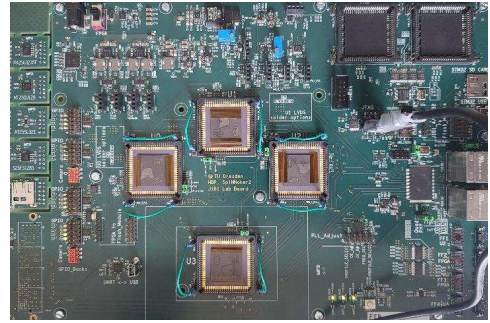}
	\caption{Testchip PCB Setup}
	\label{fig:JIB_PCB}
\end{figure}

A lab evaluation PCB has been designed, as shown in Fig.~\ref{fig:JIB_PCB}. 
It contains 6 test chip sockets, which allows for prototyping of various system integration 
aspects of the SpiNNaker2 system. 
The board is equipped with automated power supply measurement devices, 
which allows for detailed characterization of the test chip energy efficiency.   

\section{Benchmarks and Results}
\label{sec:app}
To show both, the energy efficiency of the Processing Element (PE) and the capability of hybrid digital neuromorphic of SpiNNaker2 we implement various benchmarks. This includes plain CoreMark and matrix multiplication cases and three diverse benchmark networks ranging from Synfire Chain representing the classical Spiking Neural Network (SNN) (section \ref{subsec_snn}), through the Neural Engineering Framework (NEF) representing the combined SNN/DNN approach (section \ref{subsec_nef}), to convolutional layers and fully connected layers as examples of Deep Neural Networks (DNN) (section \ref{subsec_dnn}).

\subsection{PE metrics testchip results}
\begin{figure}[htb]
	\centering
		\includegraphics[width=0.40\textwidth]{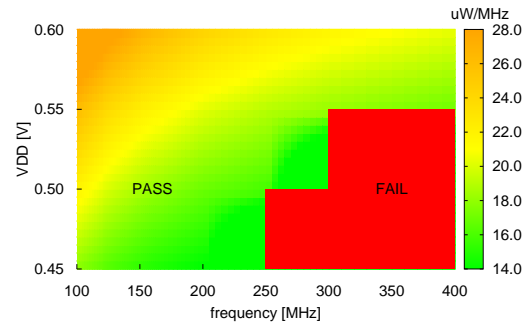}
	\caption{Measured PE CoreMark benchmark energy efficiency}
	\label{fig:PE_CoreMark_shmoo_power}
\end{figure}

\begin{figure}[htb]
	\centering
		\includegraphics[width=0.50\textwidth]{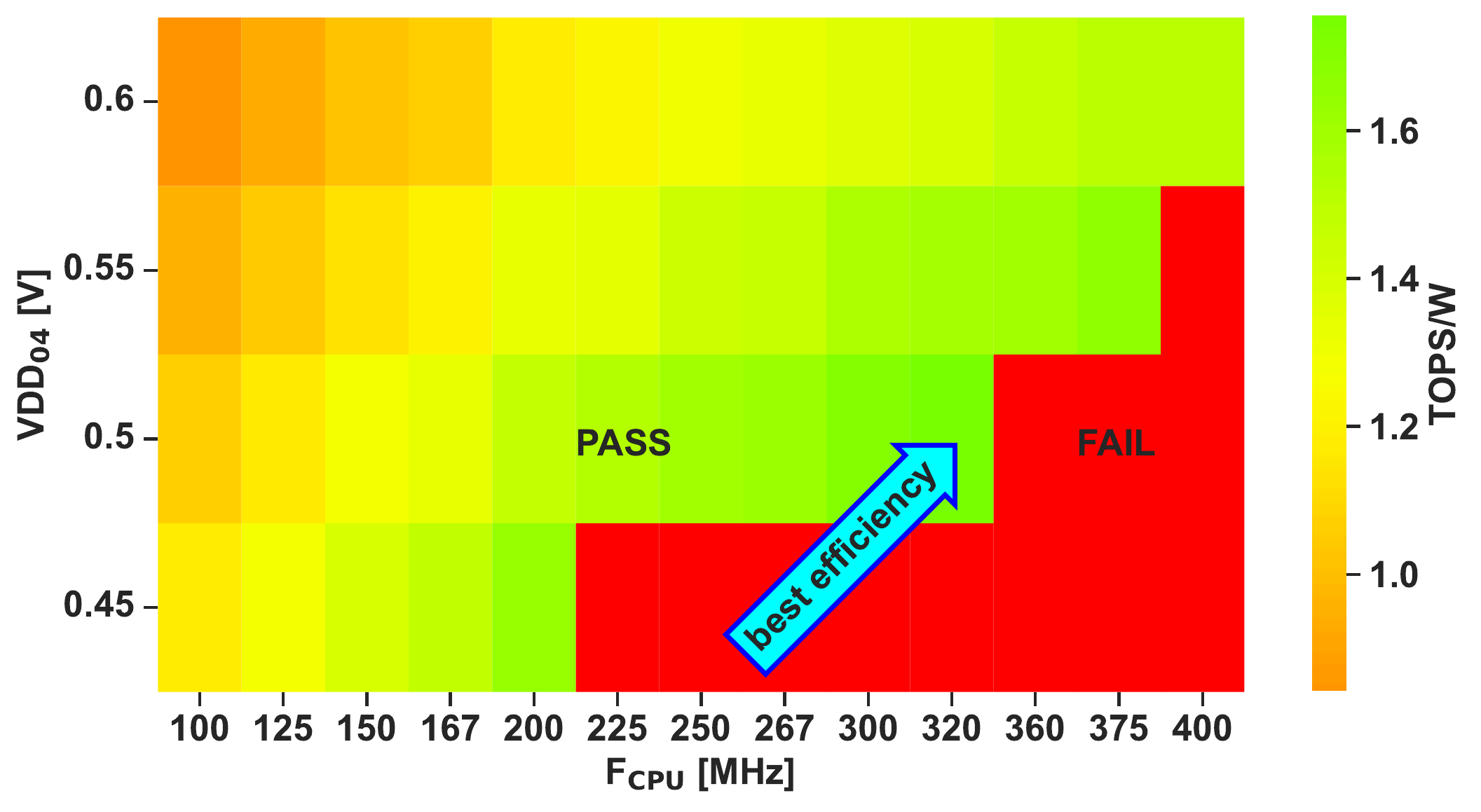}
	\caption{Measured PE matrix multiplication energy efficiency}
	\label{fig:PE_matmul_shmoo_power}
\end{figure}

Fig.~\ref{fig:PE_CoreMark_shmoo_power} shows the measured PE processor efficiency when executing the CoreMark benchmark from local SRAM. At the two DVFS PLs efficiencies of $16.68\mu \t{W}/\t{MHz}$ (at 0.50V, 200MHz) and $20.16\mu \t{W}/\t{MHz}$ (at 0.60V, 400MHz) are achieved. Fig.~\ref{fig:PE_matmul_shmoo_power} shows the measured energy efficiency when executing 8-bit matrix multiplications from local SRAM in the $16\times 4$ MAC accelerator array. At the two DVFS PLs $1.47\t{TOPS}/\t{W}$ (at 0.50V, 200MHz) and $1.51\t{TOPS}/\t{W}$ (at 0.60V, 400MHz) are achieved, respectively. For this scenario we can achieve $1.75\t{TOPS}/\t{W}$ (at 0.50V, 320MHz). Due to a hardware bug in the data transfer the overall TOPs/W are reduced by a factor of roughly $1.56$.

\subsection{Spiking Neural Network}\label{subsec_snn}

For the SNN benchmark we follow the approach of \cite{hoeppner2019dynamic}: 
each PE simulates a number of neurons and their inbound synapses. 
A timer tick triggers each time step and guarantees real time operation. 
In each time step, after the neurons are simulated, the output spikes are sent to their target PEs, 
where they are stored in a FIFO and processed in the next time step. The Performance Level (PL) of the PE,
which is a supply voltage and frequency pair (\(V_{DD},f\)), 
is adjusted according to the number of spikes in the FIFO. 
In each time step, after the simulation is finished, the PE returns to PL1 
for minimum energy consumption and switches to sleep mode, until the next timer tick wakes the PE up. 

For the testchip 3 PLs are available: PL1 with 0.5 V 100 MHz for low power operation, 
PL2 with 0.5 V 200 MHz for normal operation, and PL3 with 0.6 V 400 MHz for peak performance operation. 
Similar to \cite{hoeppner2019dynamic}, an energy model is employed to estimate the energy 
consumed in a simulation cycle:

\begin{align}
	E_{\t{cycle}}=&P_{\t{BL},i}\cdot t_\t{sp} +P_{\t{BL},1}\cdot(t_\t{sys}-t_\t{sp}) \nonumber \\
	&+e_\t{neur,i}\cdot n_\t{neur}  \nonumber \\
	&+e_\t{syn,i}\cdot n_\t{syn} \label{eq:esumcycle}
\end{align}

where \(P_{\t{BL},i}\) is the baseline power at PL \(i\), 
which means the power of a PE when it is switched on and waken up by the timer ticks, 
but after wake-up immediately goes back to sleep, without doing any computation 
related to neuron or synapse, \(t_\t{sys}\) is the length of the simulation cycle, normally 1 ms, 
\(t_\t{sp}\) is the time within the simulation cycle, during which the actual simulation is done,
\(e_\t{neur,i}\) is the energy for updating one neuron at PL \(i\), 
\(n_\t{neur}\) is the number of neurons, \(e_\t{syn,i}\) is the energy for
processing one synaptic event at PL \(i\), and \(n_\t{syn}\) is the number of synaptic events.

To extract the parameters of the energy model, 
we simulate and measure the same locally connected network as in \cite{hoeppner2019dynamic}, 
and the parameters are shown in Table~\ref{tab:npmparameters}. 

\setrowcolors{}

\begin{table}[htb]
  \begin{minipage}{0.47\textwidth}
    \centering
    \renewcommand{\arraystretch}{1.1}
    \caption{Measured parameters of energy model}
    \label{tab:npmparameters}
    \centering
    \footnotesize
    \begin{tabular}{lccc} \toprule
                                 & PL1  & PL2  & PL3 \\ 
                                 & (\SI{0.5}{V} 100 MHz) & (\SI{0.5}{V} 200 MHz) & (\SI{0.6}{V} 400 MHz) \\ \midrule
      $P_{\t{BL}}$ [\si{mW}]     & 22.38              & 29.72             & 66.44              \\
      $e_\t{neur}$    [\si{nJ}]  & 1.51               & 1.50               & 1.89               \\
      $e_\t{syn}$   [\si{nJ}]    & 0.20               & 0.20               & 0.26               \\
      \bottomrule

    \end{tabular}
  \end{minipage}
\end{table}

\begin{figure}[htb]
	\centering
		\includegraphics[width=0.47\textwidth]{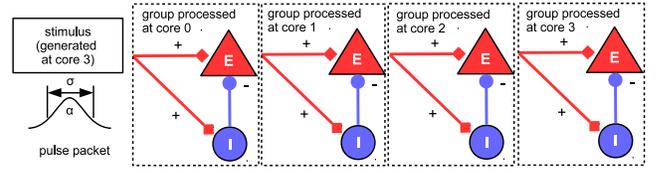}
	\caption{Synfire chain network\cite{hoeppner2019dynamic}\label{fig:synfire_chain}}
\end{figure}

Synfire chain \cite{Abeles2009} is chosen as the spiking neural network benchmark,
with the same network parameters as in \cite{hoeppner2019dynamic} and \cite{Hoeppner2017}. 
The network structure is shown in Fig.~\ref{fig:synfire_chain}. 
On each PE, there is an excitatory and an inhibitory neuron population. 
Both populations receive excitatory inputs from the previous layer (PE), 
and the excitatory population additionally receives inhibitory inputs from 
the inhibitory population on the same PE. The excitatory population has 200 neurons, 
and the inhibitory population has 50 neurons. The neurons receive a normally distributed noise current. 
The synaptic delay from the inhibitory to the excitatory population is 8 ms, 
and the synaptic delay from the excitatory population to the next layer is 10 ms. 
The last PE is connected to the first PE to form a circle. 
Each neuron has 60 presynaptic connections from the excitatory population of the last layer. 
Each excitatory neuron has 25 presynaptic connections from the inhibitory population of the same layer. 
A stimulus pulse packet is provided to PE 0 at the beginning of the simulation 
to kick start the network activity. 
The network parameters are summarized in Table \ref{tab:networks}. 
The threshold $l_\t{th}$ indicate the number of received spikes above 
which the performance level should be increased. 
These parameters were determined using the methods in \cite{hoeppner2019dynamic}.


\begin{table}[htb]
  \begin{minipage}{0.47\textwidth}
    \centering
    \renewcommand{\arraystretch}{1.1}
    \caption{Synfire chain network parameters}\label{tab:networks}
    \centering
    \footnotesize
    \begin{tabular}{lc} \toprule
      neurons per core  & 250            \\
      synapses per core & 20000        \\  
      avg.\ fan-out     & 80             \\
      $l_\t{th,1}$      & 17              \\
      $l_\t{th,2}$      & 59            \\
      \bottomrule{}
    \end{tabular}
  \end{minipage}
\end{table}

\begin{figure}[htb]
	\centering
		\includegraphics[width=0.47\textwidth]{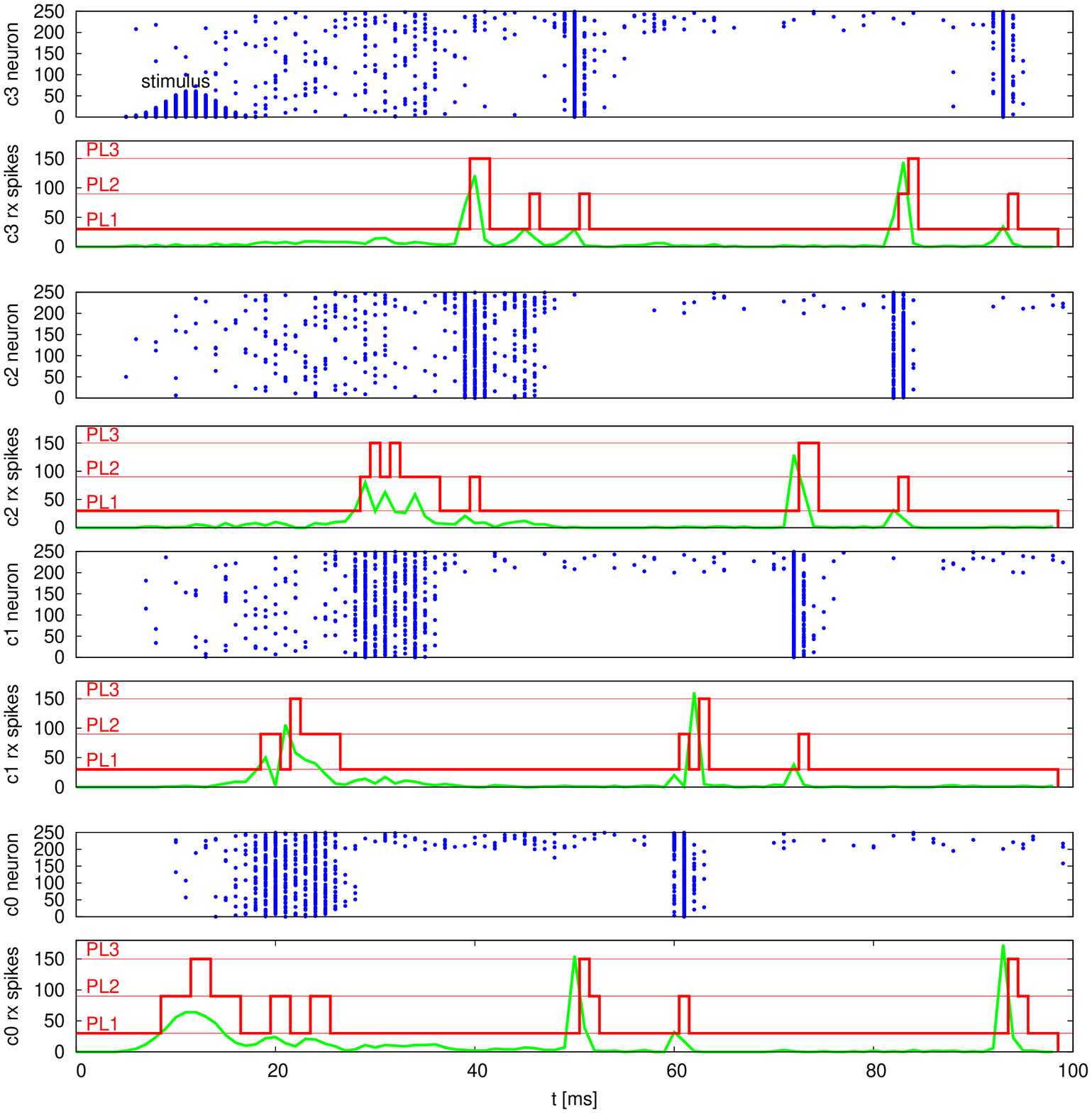}
  \caption{Synfire chain benchmark spike train, sent spikes (blue), number of received spikes per core (green) and core PL (red), plot shows zoom in to \SI{100}{ms} simulation time for illustration of fast PL changes.}\label{fig:synfire_chain_spikes_pl}.
\end{figure}

\begin{figure}[htb]
	\centering
		\includegraphics[width=0.47\textwidth]{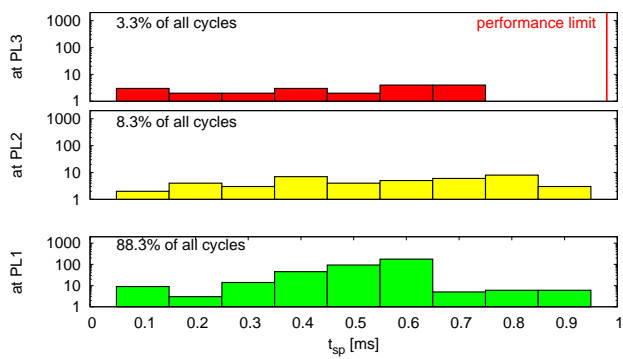}
	\caption{Histogram of simulation cycles (\SI{1}{ms}) processed at different PLs versus $t_\t{sp}$}
	\label{fig:pl_time_done_hist}
\end{figure}

The simulation and measurement show similar results as in \cite{hoeppner2019dynamic} 
and \cite{Hoeppner2017}. As shown in Fig.~\ref{fig:synfire_chain_spikes_pl}, 
the PL is increased only when it is necessary, i.e. there is more spikes to process. 
Since the network activity is very sparse for most of the time, mostly PL 1 is used 
(Fig.~\ref{fig:pl_time_done_hist}). 
Without DVFS, the system would have to be operated at PL 3 all the time. 
Power measurement is done when the PEs simulate the synfire chain with and without DVFS. The results are summarized in Table \ref{tab:dvfsresult}. With DVFS, the total power reduction is 60.4\%, with a leakage power reduction of 63.4\%. 

\begin{table}[htb]
  \begin{minipage}{0.48\textwidth}
    \centering
    \renewcommand{\arraystretch}{1.1}
    \caption{Power Measurement Results for Synfire Chain Simulation (mW)}
    \label{tab:dvfsresult}
    \centering
    \footnotesize
    \begin{tabular}{lccc} \toprule
      
                                & only PL 3 & DVFS  & reduction    \\  \midrule
      baseline power            & 66.4      & 24.3 & 63.4\% \\                       
      neuron power   & 3.3       & 2.6 & 21.2 \% \\
      synapse power    & 1.6       & 1.3 & 18.7\% \\ \midrule
      total power               & 71.3      & 28.2 & 60.4\% \\
      
      \bottomrule

    \end{tabular}
  \end{minipage}
\end{table}


\subsection{Neural Engineering Framework}\label{subsec_nef}

The Neural Engineering Framework (NEF) \cite{eliasmith2002nefbook} is a hypothesis 
about using neurons to encode scalar values or vectors. 
Computationally, it normally consists of 3 phases: encoding, neuron update and decoding. 
In the encoding process, a vector can be translated into neuron input currents. 
After the neuron update process, the neuron outputs are then translated back to a vector
in the decoding process. The encoding process is a vector-matrix multiplication, 
similar to the fully-connected layer in a Deep Neural Network. 
In the neuron update process, in the case of a spiking neuron model, 
the computation is the same as in SNN. Due to this particular computational feature, 
NEF is chosen as an example of the combined SNN/DNN approach. 

\begin{figure}[htb]
	\centering
		\includegraphics[width=0.47\textwidth]{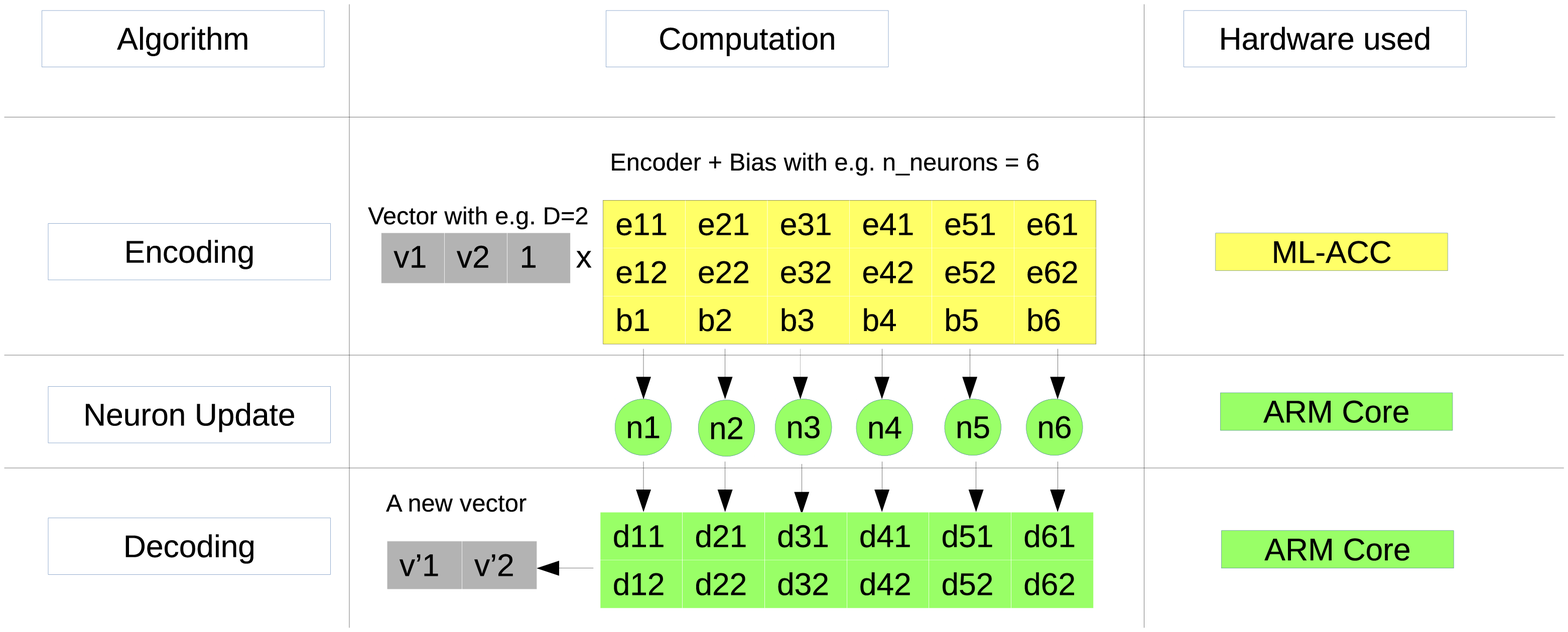}
	\caption{Computation flow of Neural Engineering Framework on test chip.}
	\label{fig:nef_mlacc}
\end{figure}

We follow the approach in \cite{mundy2015nef} to implement encoding, 
neuron update and decoding in one PE to reduce communication and computation. 
Particularly, for the test chip, since the MAC array offloads the computation of 
matrix multiplication from the ARM core, the encoding process can be executed by the MAC array. 
For spiking neurons, the decoding process is event based, so it is done in the ARM core. 
The computation is summarized in Figure \ref{fig:nef_mlacc}.

\begin{figure}[htb]
	\centering
		\includegraphics[width=0.47\textwidth]{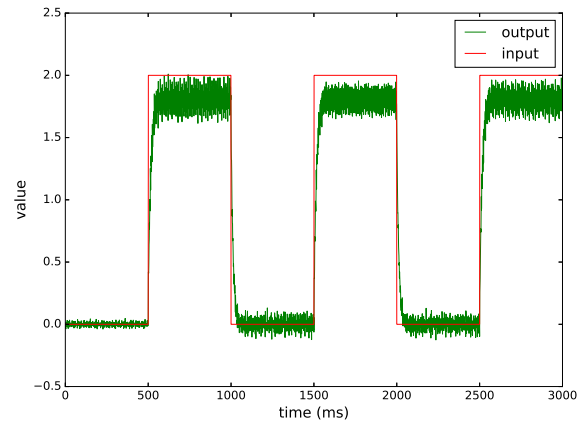}
	\caption{Communication channel with 512 neurons and 1 dimension. The decoded output of the neuron population tries to resemble the input.}
	\label{fig:nef_sim}
\end{figure}

To demonstrate the functionality of the implementation, 
we show the communication channel example, 
where the decoded output of a neuron population tries to resemble the input vector. 
Figure \ref{fig:nef_sim} shows the result of the simulation. 
The neuron population consists of 512 neurons and represents 1 dimension. 

The communication channel is simulated on the 8 PEs of the test chip and the dynamic power 
(i.e. total power subtracted by baseline power) is measured. The test chip runs with 0.5V 200 MHz. 

\begin{figure}[htb]
	\centering
		\includegraphics[width=0.5\textwidth]{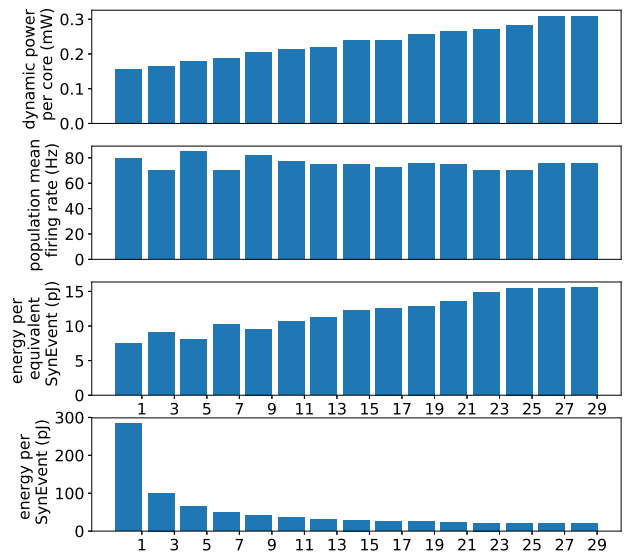}
	\caption{Dynamic power per core, population mean firing rate, energy per synaptic event and energy per equivalent synaptic event, with 512 neurons.
  }
	\label{fig:nef_meas}
\end{figure}

Dividing the dynamic energy of one time step by the number of synaptic events in one time step,
we can get the energy per synaptic event as an energy metric which is commonly used 
in the neuromorphic literature. For the following calculations, 
we consider two neuron populations each with \(N\) neurons and \(D\) dimensions. 
To calculate the number of synaptic events, two approaches are considered. 
The first approach is the equivalent synaptic event similar to Braindrop \cite{braindrop}. 
Here, we consider the number of synaptic events it would be if the connection matrix were not factorized, 
i.e. \(N N\) connections between two populations, and each spike causing \(N\) synaptic operations. 
The second approach is to consider the hardware operations related to the synaptic events, 
i.e. the \(N D\) MAC operations done by the MAC array and the \(D\) ADD operations 
for each neuron that has spiked in a time step. 
Assuming \(M\) neurons have spiked in a time step, 
the number of synaptic operations is \(N D + M D\).  
The results of both approaches are shown in Fig. \ref{fig:nef_meas}. 
In the first approach, the energy per equivalent synaptic operation is ca. 10 pJ, 
surpassing Loihi, which has reported 24 pJ per synaptic operation \cite{Davies2018}. 
In the second approach, for higher dimensions, the energy per synaptic operation reaches 20 pJ,
approaches and slightly surpasses Loihi. 



\subsection{Deep Neural Network}\label{subsec_dnn}
\Com{Florian}

\begin{figure}[htb]
	\centering
		\includegraphics[width=0.50\textwidth]{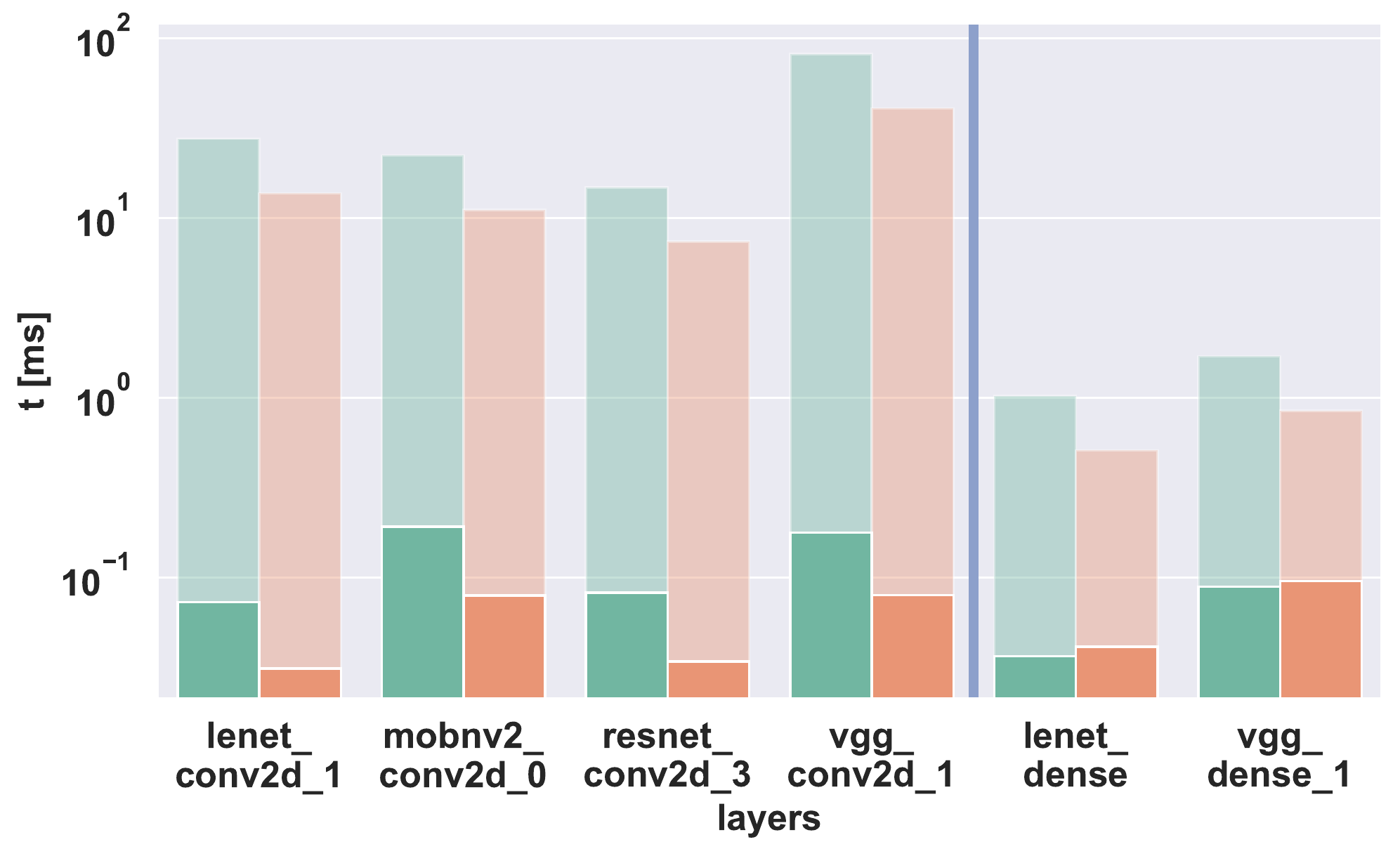}
	\caption{Comparison of run time of a subpart of DNN layers between an ARMNN implementation and the accelerator. For each layer we show the frequency-voltage pairs (200MHz, 0.5V) in green and (400MHz, 0.6V) in orange. The lightly colored bar plots are representing the time the ARMNN implementation needs and the saturated bars exhibit the run time of the accelerator. The time axis is set to a log scale. In the stated scenarios we can reach a speed up between 116 and 610 for convolutional layers and a speed up between 9 and 28 for matrix multiplication.}
	\label{fig:bar_t_ms_armnn}
\end{figure}

\begin{figure}[htb]
	\centering
		\includegraphics[width=0.50\textwidth]{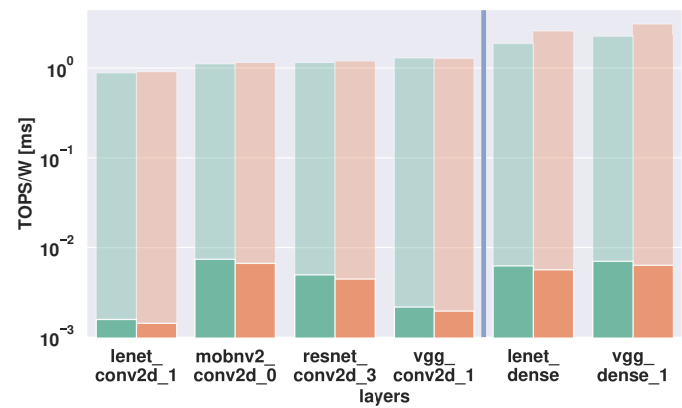}
	\caption{Comparison of energy efficiency of a subpart of DNN layers between an ARMNN implementation and the accelerator. In comparison to Fig. \ref{fig:bar_t_ms_armnn} the lightly colored plots are now representing the accelerator while the saturated bars show the efficiency of the accelerator. We can reach an efficiency increase by a factor between 148 and  652 for convolutional layers and by a factor between 297 and 482 for fully-connected layers with the chosen sublayers.}
	\label{fig:bar_topsw}
\end{figure}

For benchmarking the 2D cross-correlation and matrix multiplication accelerator we select individual layers from Lenet \cite{lecun1998}, VGG-16 \cite{simonyan2014very}, ResNet-50 \cite{He_2016_CVPR} and MobileNetV2 \cite{sandler2018} and compare the results with a comparable execution with ARMNN\cite{lai2018armnn} on the ARM4F core. For a study of complete DNNs, please refer to the simulation study by \cite{kelber2020mapping}. We divide the layers to fit into the 128kByte SRAM per PE and optimize to utilize each MAC cell of the accelerator as much as possible. The data is send into each of the 8 PEs and then processed 100000 times in a loop. For time measurements we use the ARM core SysTick timer. Because of the possibility to control PE clock frequency and supply voltage we sweep over both while reading out shunt voltage of each power lane with a sampling rate of 100ms. Fig. \ref{fig:bar_t_ms_armnn} presents the speed up of the chosen layers and Fig. \ref{fig:bar_topsw} the gained energy efficiency respectively.

- should I write up some meta data ?
- As single runs are very short, we measures in the loop for 1 or 2 seconds. Timer measurements are used for throughput calculation 
- shmoo show TOPS/W on the right bar
- shmoo conv if better values

\subsection{Results Summary}


\section{Conclusion}
Outlook/Conclusion:
paper shows of the range of possible network types that SpiNNaker2 can be used for. 
Especially hybrid neural networks, i.e. various crossovers between SNNs and DNNs, 
can benefit from the SpiNNaker2 architecture. Possibilities of combining MACs with spiking neurons:
-as shown, for multi-bit synaptic current input to spiking neurons
-Dynamically sparse DNNs, i.e. where information is transmitted only on some form of change, not on a frame-by-frame basis.
-a learning-to-learn loop where the inner network is a spiking one and the outer network is a DNN optimizing some target function by modifying the inner spiking network
-multi-scale modelling, DNNs realize rate-based or mesoscopic networks, interface with spiking networks

go through metrics/results, compare to other hardware

state-of-the-art ML:
- Sticker-T \cite{yue20197} achieves 50 TOPs/W 4-bit

Industry overview: \cite{rueckert2020digital}

hybrid: \cite{pei2019towards}
Tianjic\cite{deng2020tianjic} reports 1278 GOPs/W DNN and 649 GSOPS/W (synaptic operations).

Other architectures to compare:
- TrueNorth SNN + eventuell Joule per inference for DNN tasks
- Loihi
- BrainDrop for equivalent SynOp?
- key-word spotting paper, which compares different approaches.

latest BrainScales-2 system has option for analog vector-matrix multiplication and 
HAGEN-mode HICANN-DLS \cite{schemmel2020accelerated,weis2020inference}.
  - 5-bit input, 6-bit weight, 8-bit neuron activation
  - 12 mJ per image (conv network) 3 mJ for smaller dense network

  - SNN: \cite{cramer2020training} 4uJ per image. Simple MLP, with latency code. They try spike sparsity by adding a term to the loss function

Interesting comparison of SNN and DNN by UC Waterloo for same network architecture:
\cite{blouw2020event}
They actually use a conv operation in the input layer.

Maybe add some paper from UNIBI: Christoph Ostrau \cite{ostrau2020benchmarking,ostrau2019comparing} 
benchmarking of SNN hardware + CPU + GPU

What actually matters: energy to solution. However no clear benchmarks defined \cite{davies2019benchmarks}
GPUs are also a candidate: see e.g.: \cite{knight2018gpus}: down to 0.3 uJ / synaptic event on TX2.
A draft comparison table can be found in file: "CompHybridNeuromorph.ods"

\label{sec:conclusion}



\begin{IEEEbiography}[{\includegraphics[width=1in,height=1.25in,clip,keepaspectratio]{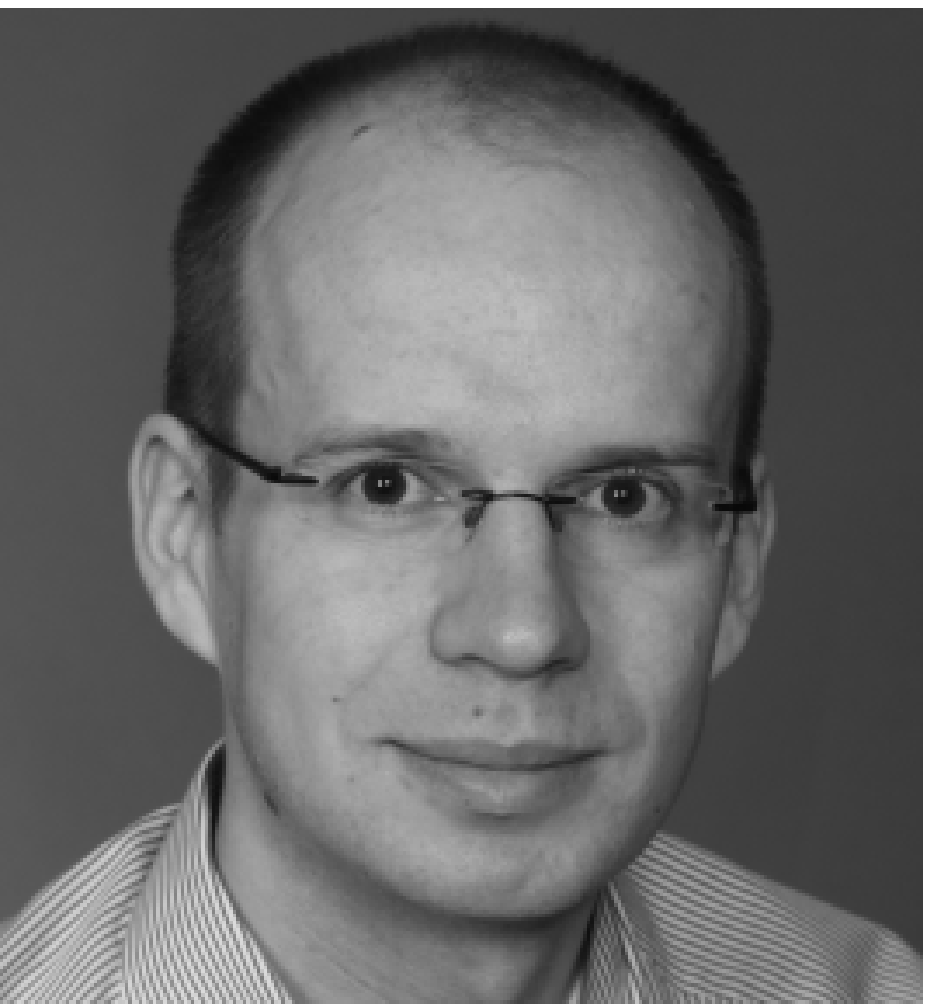}}]{Sebastian Höppner}
is a Research Group Leader and Lecturer with the Chair of Highly-Parallel VLSI-Systems and Neuromorphic Circuits. He received the Dipl.-Ing. (M.Sc.) in Electrical Engineering in 2008 and his Ph.D. in 2013 (received Barkhausen Award), both Technische Universität Dresden, Germany. His research interests include circuits for low-power systems-on-chip in advanced technology nodes, with special focus on clocking, data transmission and power management. He has experience in designing full-custom circuits for multi-processor systems-on-chip (MPSoCs), like ADPLLs, register files and high-speed on-chip and off-chip links, in academic and industrial research projects. He has been managing the full-custom circuit design and SoC integration for more than 12 MPSoC chips in 65nm, 28nm and 22nm CMOS technology. Currently he leads the chip design of the SpiNNaker2 neuromorphic computing system within the Human Brain Project(HBP). He is author or co-author of more than 56 publications and 10 patents (5 issued, 5 pending) in the above fields.
\end{IEEEbiography}

\begin{IEEEbiography}[{\includegraphics[width=1in,height=1.25in,clip,keepaspectratio]{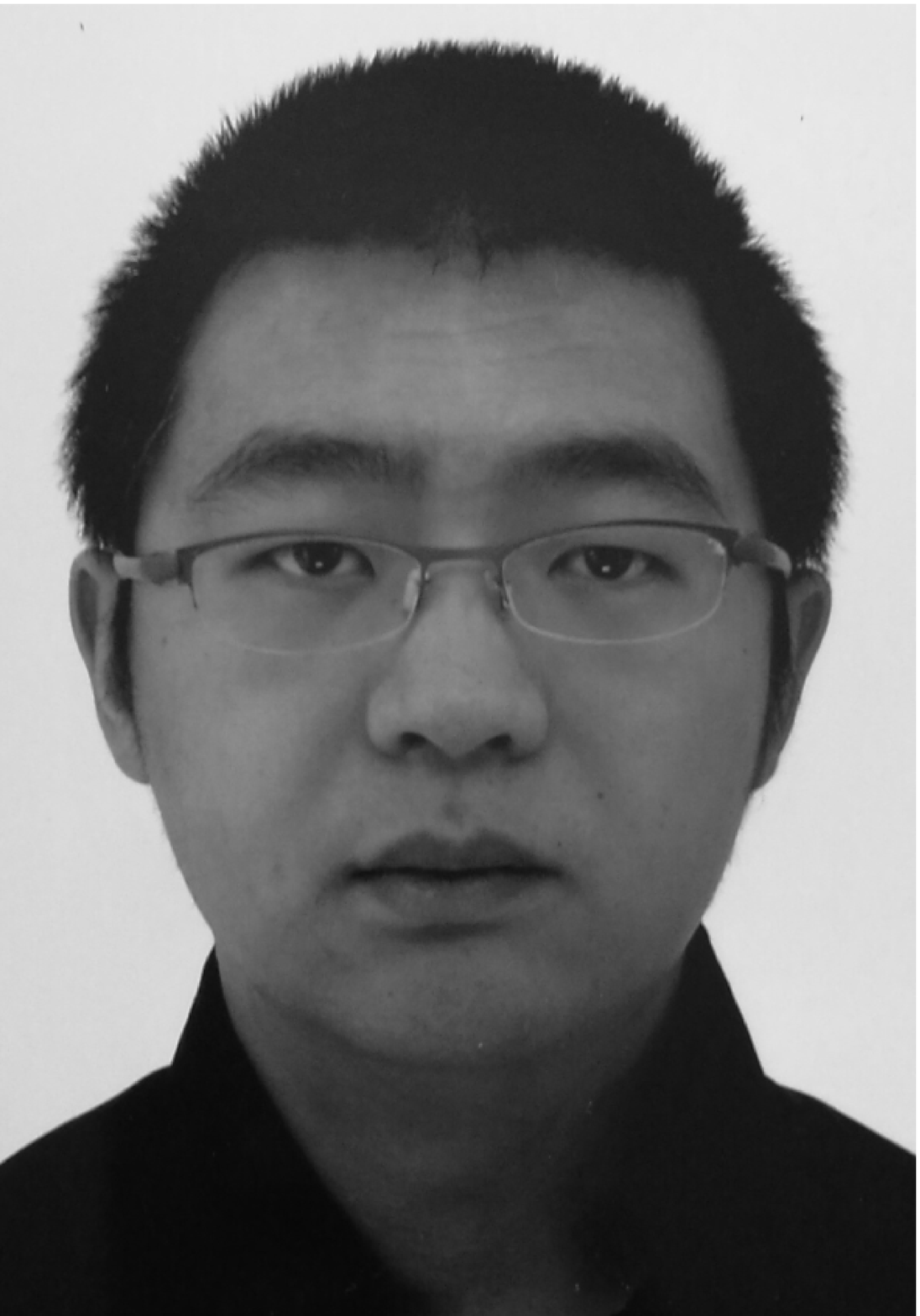}}]{Yexin Yan}
received the Dipl.-Ing. (M.Sc.) in Electrical Engineering from Technische Universität Dresden, Germany, in 2016. He is currently pursuing the Ph.D. at the Chair of Highly-Parallel VLSI-Systems and Neuromorphic Circuits at Technische Universität Dresden. His research interests include hardware-software co-design for low power neuromorphic applications.
\end{IEEEbiography}

\begin{IEEEbiography}[{\includegraphics[width=1in,height=1.25in,clip,keepaspectratio]{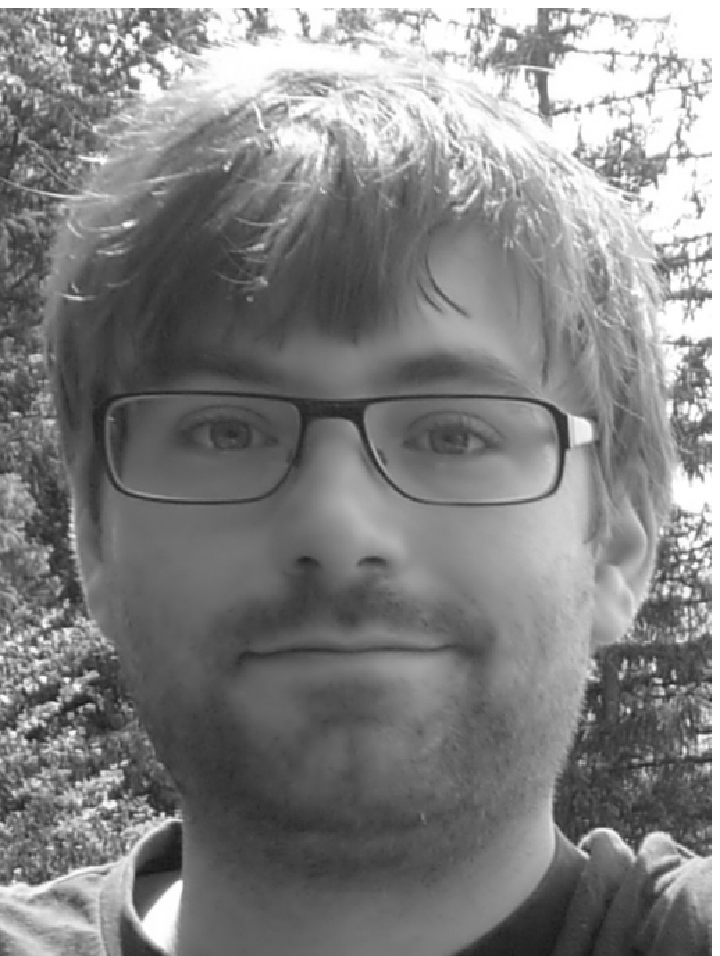}}]{Bernhard Vogginger}
received the diploma in physics from the University of Heidelberg, Heidelberg, Germany, in 2010. Currently, he is a research associate at the Chair of Highly-Parallel VLSI-Systems and Neuromorphic Circuits at Technische Universität Dresden, Germany, where he is pursuing a PhD under the supervision of Prof. Christian Mayr. His research interests include neuromorphic engineering, neural computation and deep learning.
\end{IEEEbiography}

\begin{IEEEbiography}[{\includegraphics[width=1in,height=1.25in,clip,keepaspectratio]{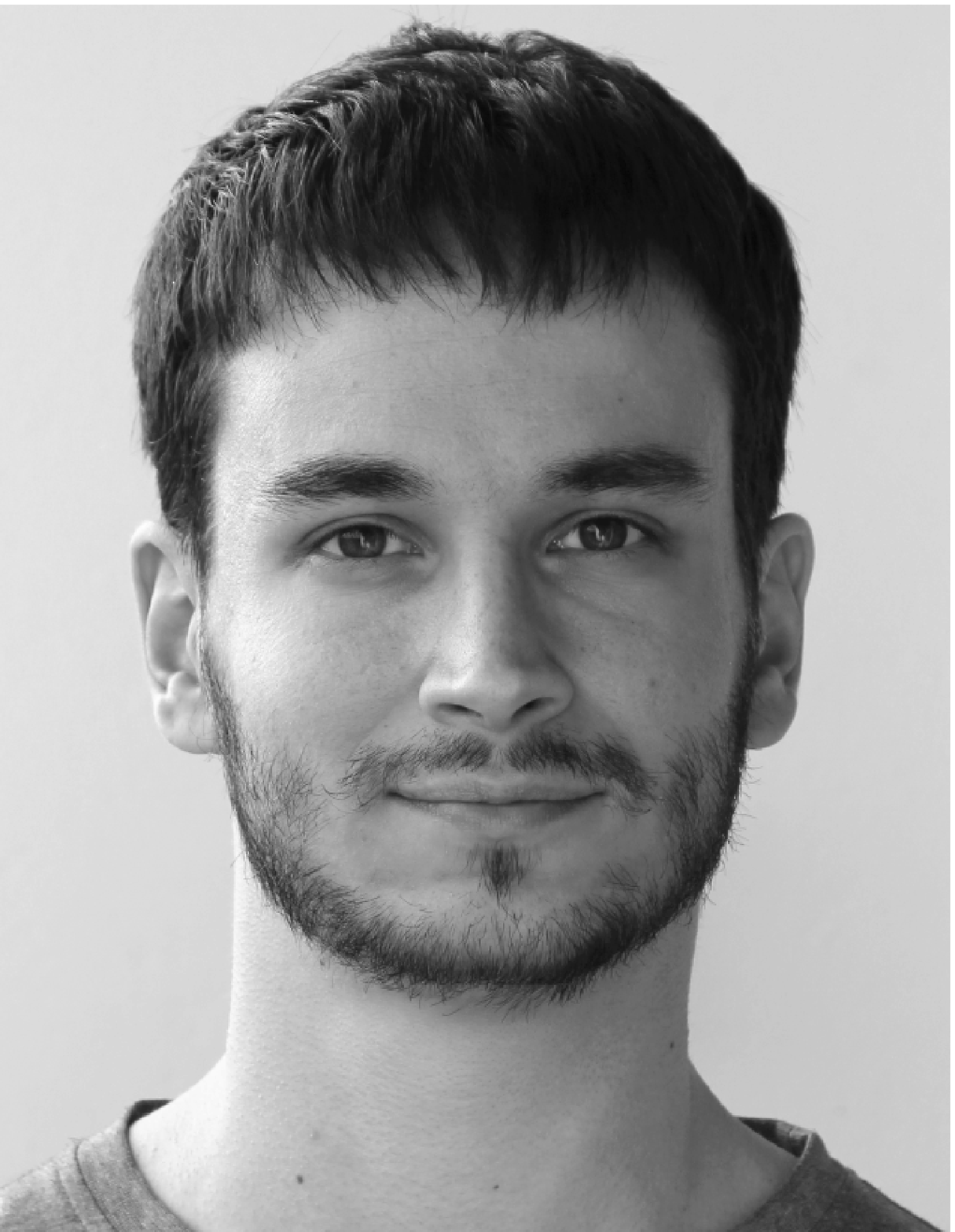}}]{Chen Liu}
received the Dipl.-Ing. (M.Sc.) in 
\end{IEEEbiography}

\begin{IEEEbiography}[{\includegraphics[width=1in,height=1.25in,clip,keepaspectratio]{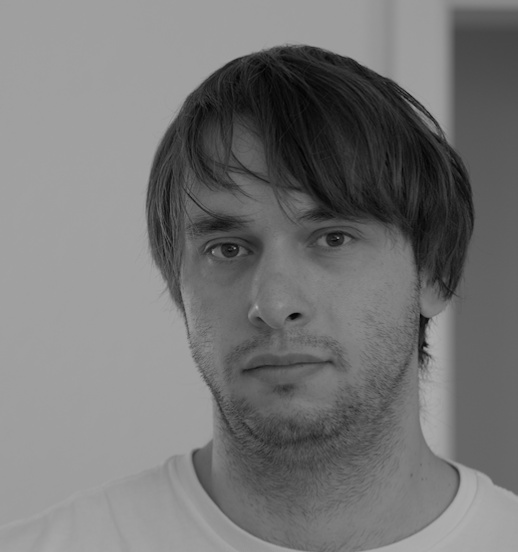}}]{Marco Stolba}
is a PhD student with the Chair of Highly-Parallel VLSI-Systems and Neuromorphic Circuits. He received his MSc degree in Nanoelectronic Systems in 2016 from TU Dresden. His research interests include network on chip (NoC), MPSoC architectures, digital design and verification.
\end{IEEEbiography}

\begin{IEEEbiography}[{\includegraphics[width=1in,height=1.25in,clip,keepaspectratio]{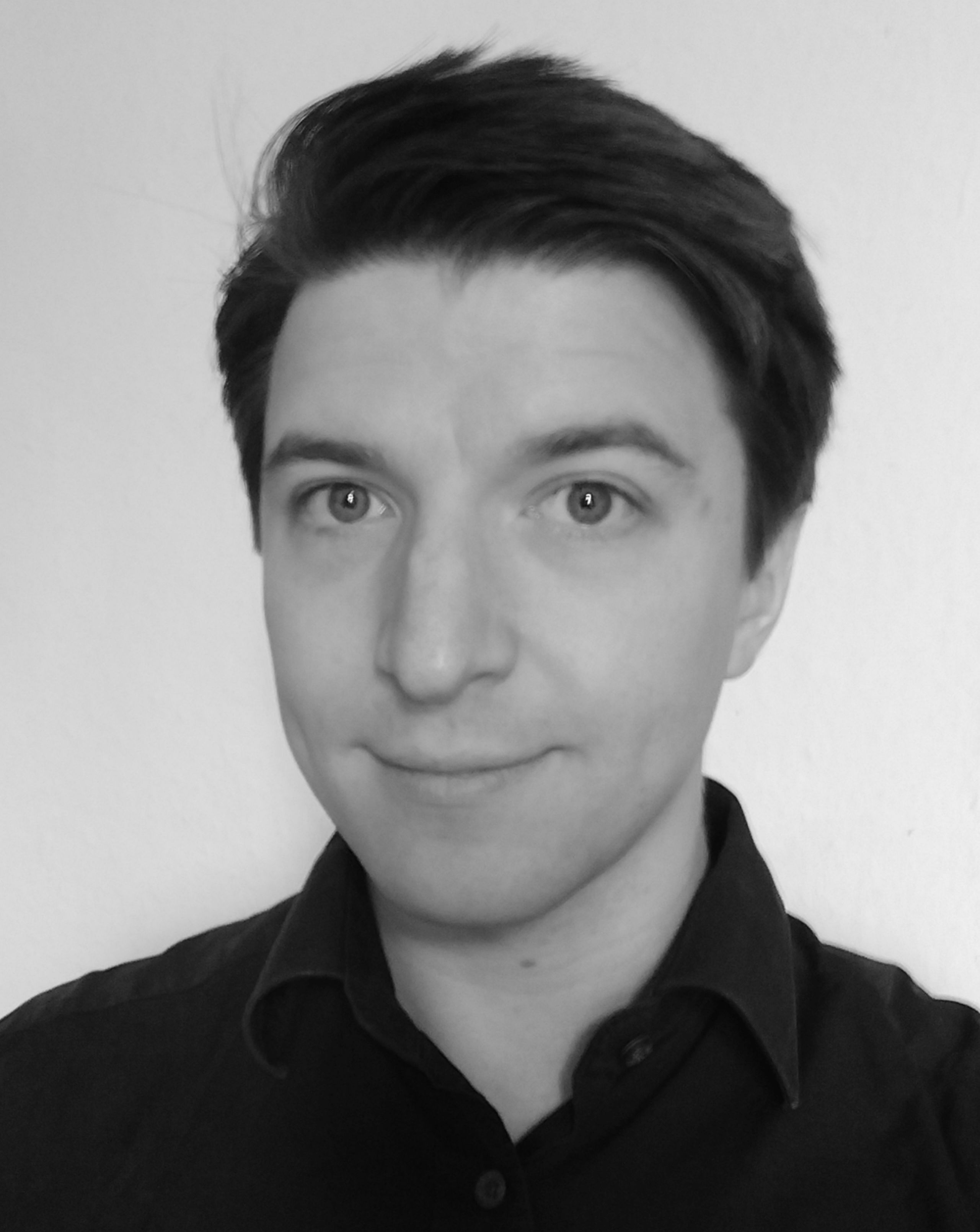}}]{Florian Kelber}
is a PhD student at the Chair for Highly-Parallel VLSI-Systems and Neuromorphic circuits at Technische Universit\"at Dresden, Germany. He received the Dipl.-Ing. (M.Sc.) in Information Systems Engineering at degree from Technische Universit\"at Dresden, Germany. His research focus is the design of digital low-power accelerators to mitigate bottlenecks in multi-processor system-on-chip systems and mapping of load-balanced high level algorithms on highly parallel architectures. 
\end{IEEEbiography}

\begin{IEEEbiography}[{\includegraphics[width=1in,height=1.25in,clip,keepaspectratio]{./bio_photos/photo_dixius}}]{Andreas Dixius}
received the Dipl.-Ing. (M.Sc.) in Information Systems Engineering at Technische Universität Dresden, Germany in 2014. Since 2014 he has been a research assistant at the Chair of Highly-Parallel VLSI-Systems and Neuromorphic Circuits, Technische Universität Dresden, Germany. His research interests include on-chip timing-detection.
\end{IEEEbiography}

\begin{IEEEbiography}[{\includegraphics[width=1in,height=1.25in,clip,keepaspectratio]{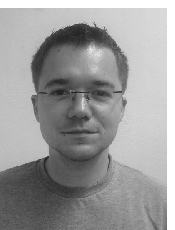}}]{Stefan Scholze}
received the Dipl.-Ing. (M.Sc.) in Information Systems Engineering from Technische Universität Dresden, Germany in 2007. Since 2007, he has been a research assistant at the Chair of Highly-Parallel VLSI-Systems and Neuromorphic Circuits, Technische Universität Dresden, Germany. His research interests include design and implementation of low-latency communication channels and systems.
\end{IEEEbiography}

\begin{IEEEbiography}[{\includegraphics[width=1in,height=1.25in,clip,keepaspectratio]{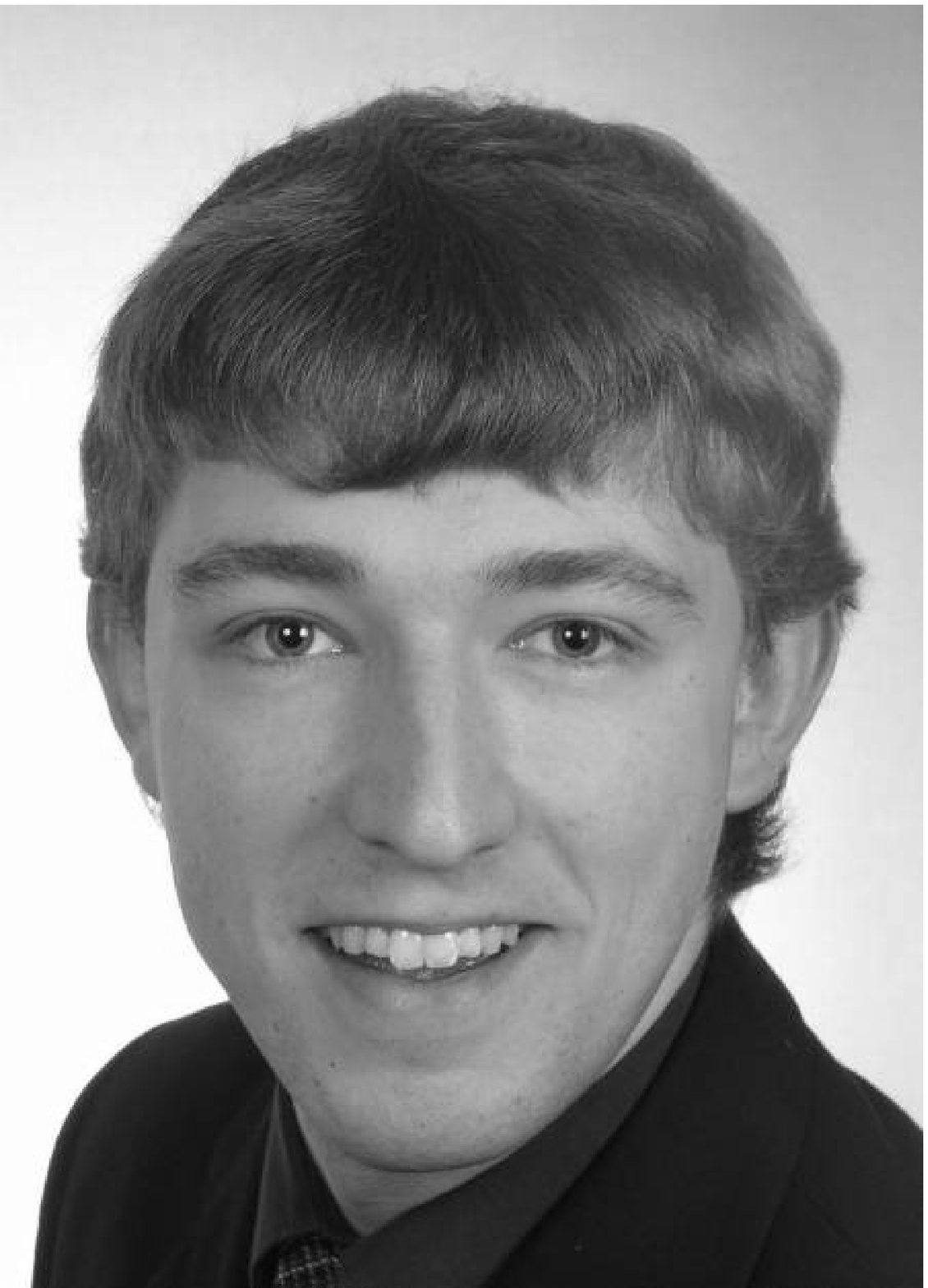}}]{Johannes Partzsch}
obtained his M.Sc.\  in Electrical Engineering in 2007 and his PhD in 2014, both from Technische Universität Dresden. He is currently a Research Group Leader at the Chair of Highly-Parallel VLSI-Systems and Neuromorphic Circuits, Technische Universität Dresden, Germany. His research interests include neuromorphic systems design, topological analysis of neural networks and technical application of bio-inspired systems. He is author or co-author of more than 45 publications.
\end{IEEEbiography}

\begin{IEEEbiography}[{\includegraphics[width=1in,height=1.25in,clip,keepaspectratio]{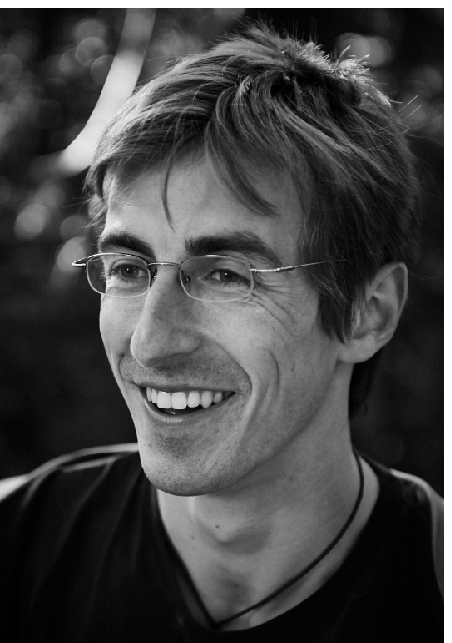}}]{Felix Neumärker}
received the Dipl.-Ing. (M.Sc.) in Electrical Engineering from Technische Universität Dresden, Germany, in 2015.
He is currently working as research associate with the Chair of Highly-Parallel VLSI-Systems and Neuromorphic Circuits at Technische Universität Dresden.
His research interests include software and circuit design and for MPSoCs with special focus on neuromorphic computing.
\end{IEEEbiography}

\begin{IEEEbiography}[{\includegraphics[width=1in,height=1.25in,clip,keepaspectratio]{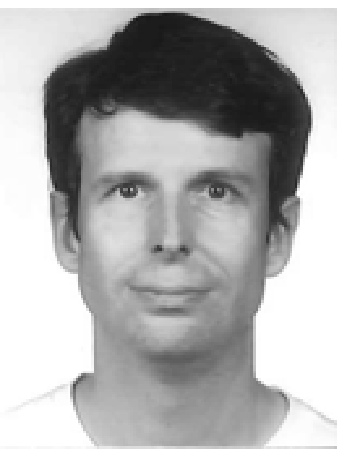}}]{Stephan Hartmann}
received the Dipl.-Ing. (M.Sc.) in Electrical Engineering from Technische Universität Dresden, Germany, in 2007. He is
currently working as research associate with the Chair of Highly-Parallel VLSI-Systems and Neuromorphic Circuits at Technische
Universität Dresden. His research interests include circuit design with special focus on FPGA.
\end{IEEEbiography}

\begin{IEEEbiography}[{\includegraphics[width=1in,height=1.25in,clip,keepaspectratio]{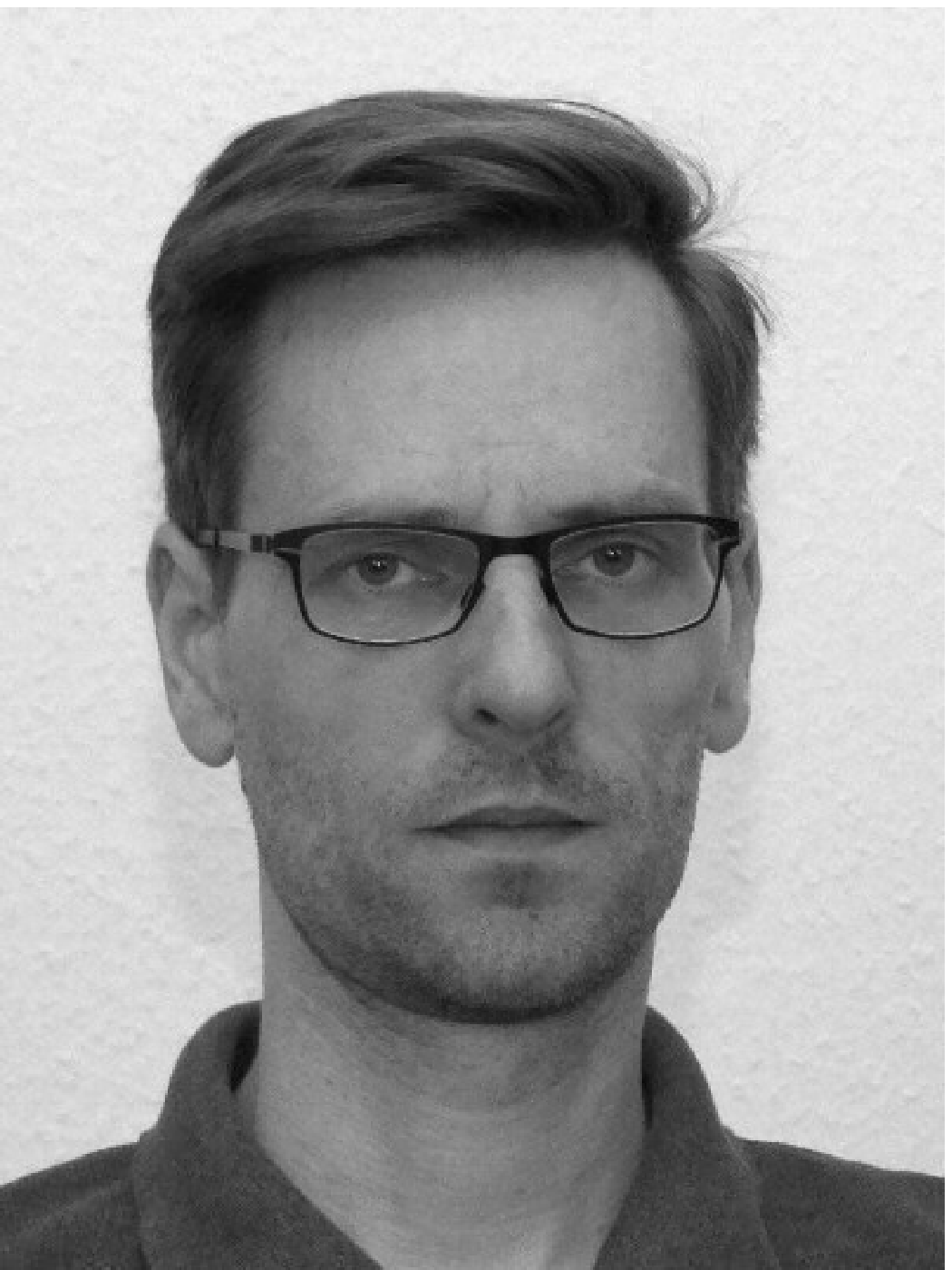}}]{Stefan Schiefer}
received the Dipl.-Ing. (M.Sc.) in Electrical Engineering from Technische Universität Dresden, Germany in 2008.
He is currently working as research associate with the Chair of Highly-Parallel VLSI-Systems and Neuromorphic Circuits at Technische Universität Dresden.
His research interests include full system signal and power integrity as well as regulator behavioural modelling including non-linearities.
\end{IEEEbiography}

\begin{IEEEbiography}[{\includegraphics[width=1in,height=1.25in,clip,keepaspectratio]{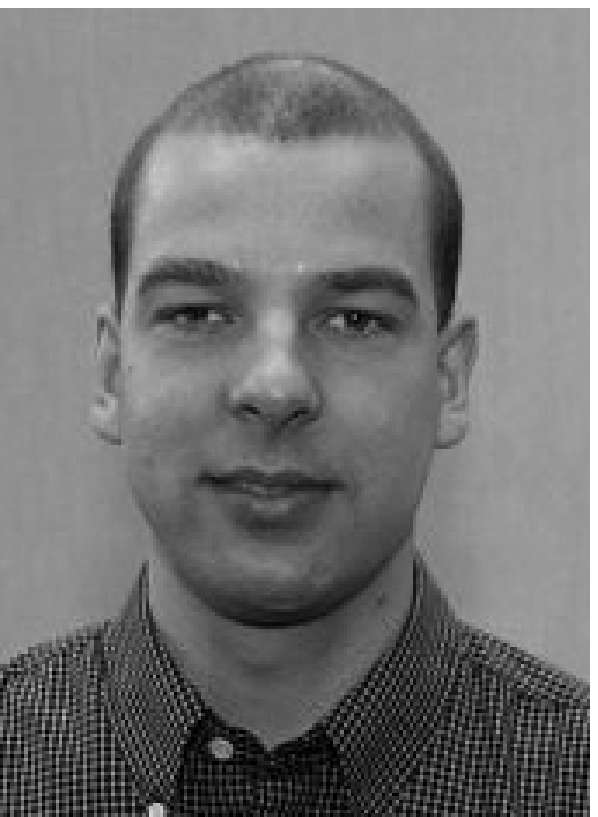}}]{Georg Ellguth}
received the Dipl.-Ing. (M.Sc.) in Electrical Engineering from Technische Universität Dresden, Germany in 2004. Since 2004, he has been a research assistant with the Chair of Highly-Parallel VLSI-Systems and Neuromorphic Circuits at Technische Universität Dresden. His research interests include low-power implementation techniques in multi-processor system-on-chip.
\end{IEEEbiography}


\begin{IEEEbiography}[{\includegraphics[width=1in,height=1.25in,clip,keepaspectratio]{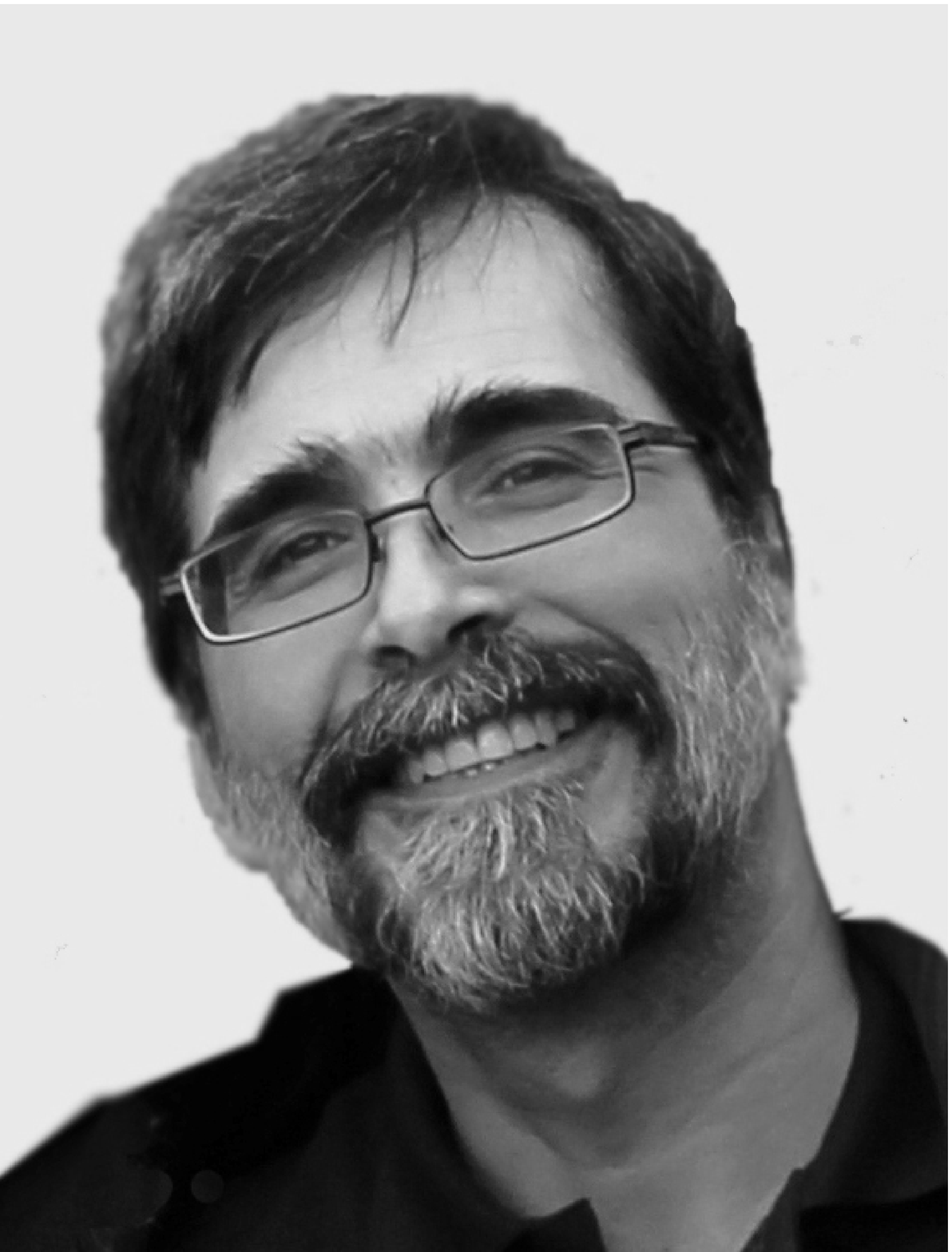}}]{Luis A Plana}
(M'97-SM'07) received the Ingeniero Electrónico (Cum Laude) degree from Universidad Simón Bolívar, Venezuela, and the PhD degree in computer science from Columbia University, USA.\@ He was with Universidad Politécnica, Venezuela, for over 20 years, where he was Professor of Electronic Engineering. Currently, he is a Research Fellow in the School of Computer Science, University of Manchester, UK.
\end{IEEEbiography}

\begin{IEEEbiography}[{\includegraphics[width=1in,height=1.25in,clip,keepaspectratio]{./bio_photos/photo_dixius}}]{Mantas Mikaitis}
received the Dipl.-Ing. (M.Sc.) in 
\end{IEEEbiography}

\begin{IEEEbiography}[{\includegraphics[width=1in,height=1.25in,clip,keepaspectratio]{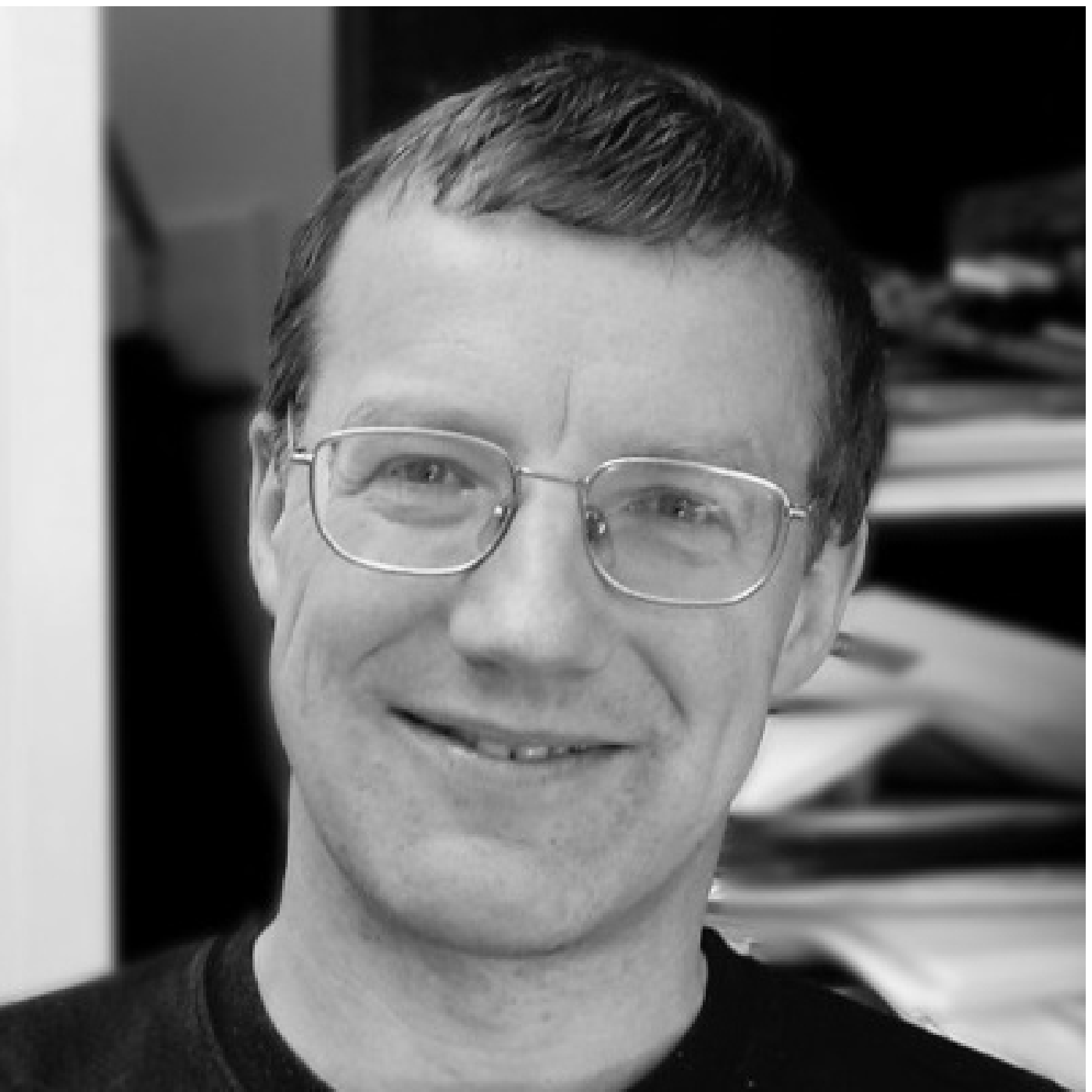}}]{Jim Garside}
received the Ph.D. degree in computer science from The University of Manchester, Manchester, U.K., in 1987, for work in signal processing architecture.  Post-doctoral work on parallel processing systems based on Inmos Transputers was followed by a spell in industry writing air traffic control software.  Returning to academia gave an opportunity for integrated circuit design work, dominated by design and construction work on asynchronous microprocessors in the 1990s. He has been involved with dynamic hardware compilation, GALS interconnection, and the development of the hardware and software of the SpiNNaker neural network simulator.
\end{IEEEbiography}

\begin{IEEEbiography}[{\includegraphics[width=1in,height=1.25in,clip,keepaspectratio]{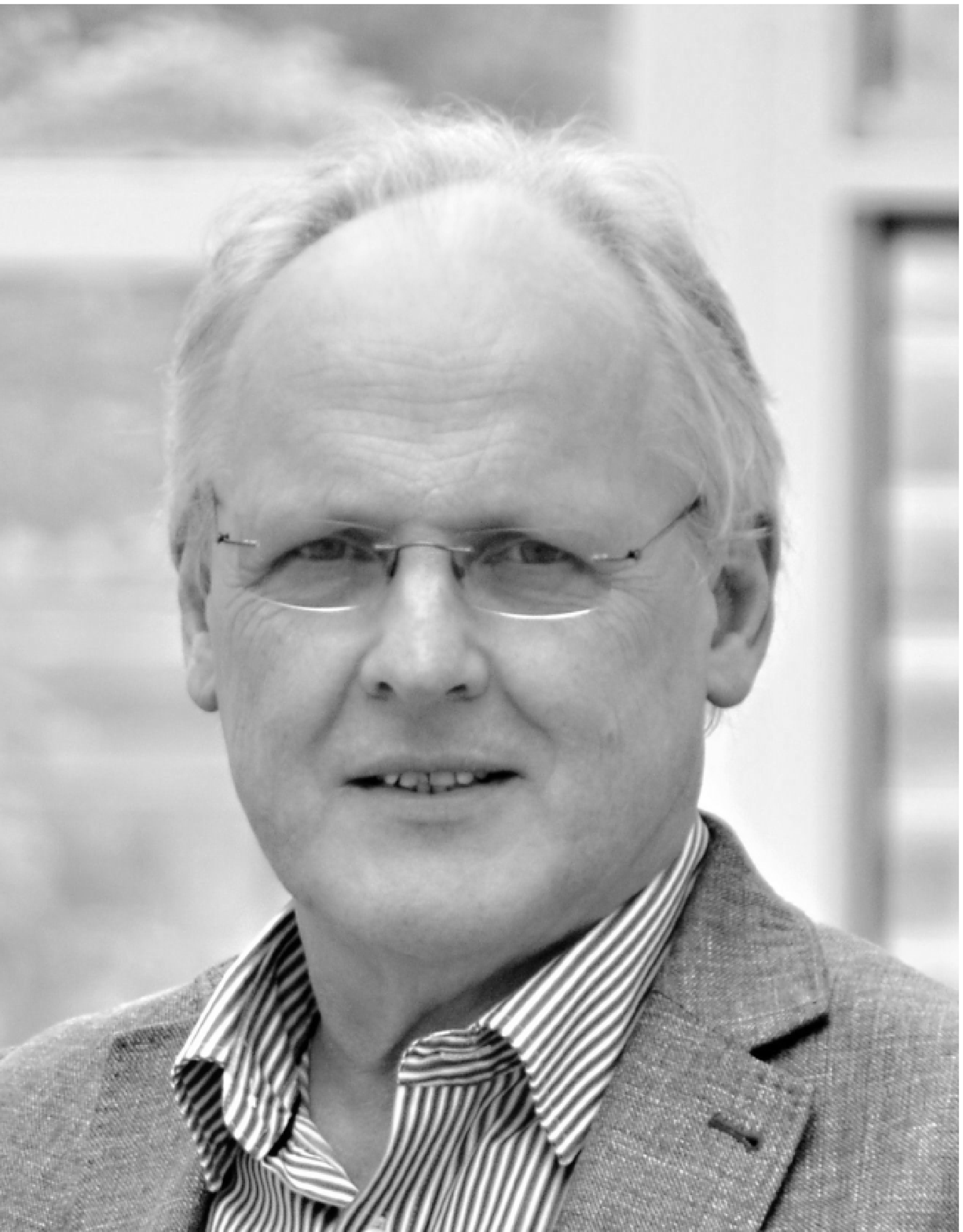}}]{Steve Furber}
CBE FRS FREng is ICL Professor of Computer Engineering in the School of Computer Science at the University of Manchester, UK. After completing a BA in mathematics and a PhD in aerodynamics at the University of Cambridge, UK, he spent the 1980s at Acorn Computers, where he was a principal designer of the BBC Microcomputer and the ARM 32-bit RISC microprocessor. Over 120 billion variants of the ARM processor have since been manufactured, powering much of the world's mobile and embedded computing. He moved to the ICL Chair at Manchester in 1990 where he leads research into asynchronous and low-power systems and, more recently, neural systems engineering, where the SpiNNaker project is delivering a computer incorporating a million ARM processors optimised for brain modelling applications.
\end{IEEEbiography}

\begin{IEEEbiography}[{\includegraphics[width=1in,height=1.25in,clip,keepaspectratio]{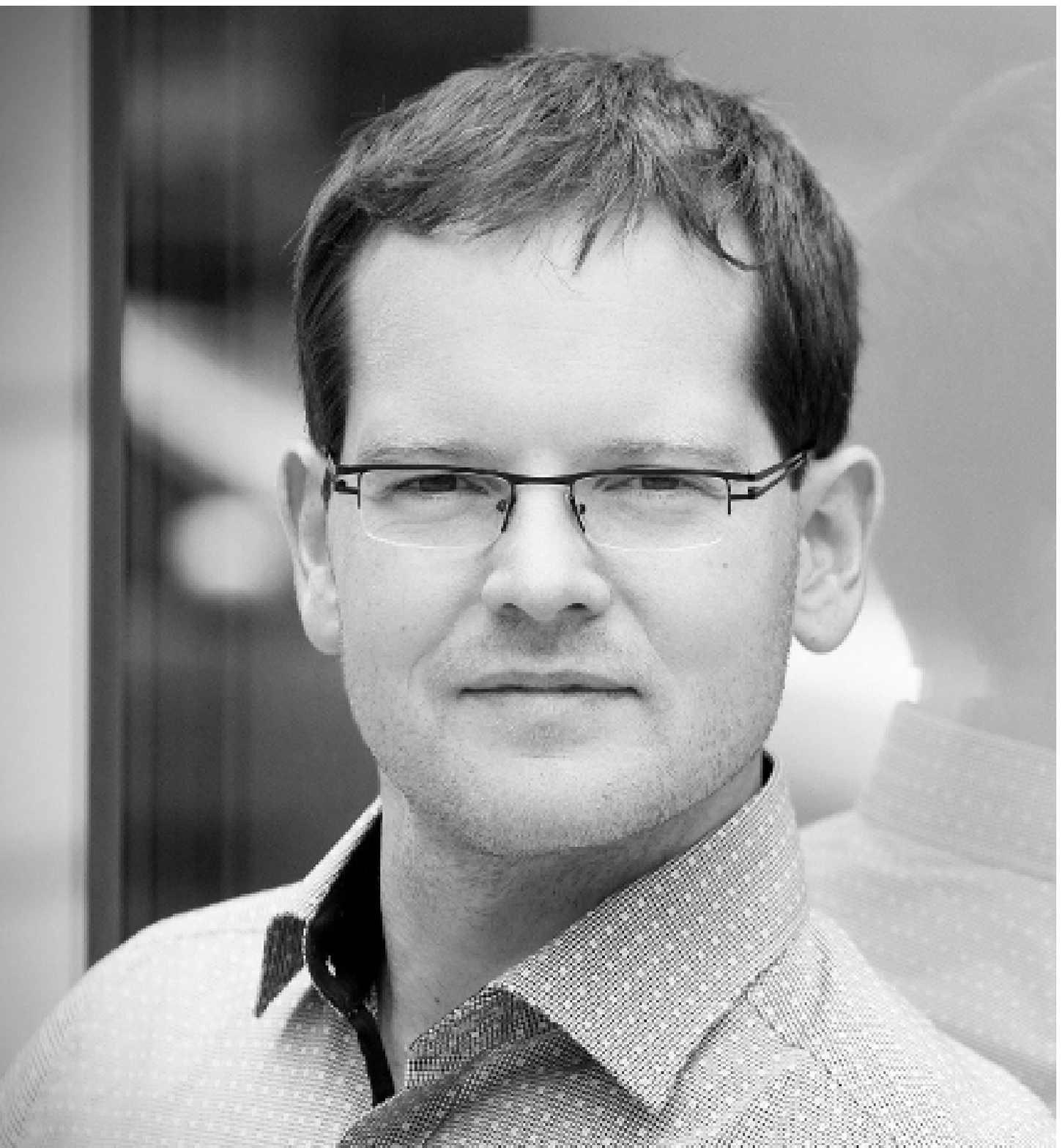}}]{Christian Mayr}
is a Professor of Electrical Engineering at TU Dresden. He received the Dipl.-Ing. (M.Sc.) in Electrical Engineering in 2003, his PhD in 2008 and Habilitation in 2012, all three from Technische Universität Dresden, Germany.
From 2003 to 2013, he has been with Technische Universität Dresden, with a secondment to Infineon (2004-2006). From 2013 to 2015, he did a Postdoc at the Institute of Neuroinformatics, University of Zurich and ETH Zurich, Switzerland. Since 2015, he is head of the Chair of Highly-Parallel VLSI-Systems and Neuromorphic Circuits at Technische Universität Dresden. His research interests include bio-inspired circuits, brain-machine interfaces, AD converters and general mixed-signal VLSI-design. He is author/co-author of over 80 publications and holds 4 patents. He has acted as editor/reviewer for various IEEE and Elsevier journals. His work has received several awards.
\end{IEEEbiography}

\end{document}